\documentclass[amsmath,amssymb,aps,pre,onecolumn,notitlepage]{revtex4-1}

\usepackage{graphicx}
\usepackage{dcolumn}
\usepackage{bm}

\usepackage{titlesec}
\usepackage{hyperref}

\usepackage{color}

\newcommand{\condsub}[1]{\Phi_{#1}}
\newcommand{\condij}[0]{\Phi_{ij}}
\newcommand{\condji}[0]{\Phi_{ji}}

\newcommand{\coups}[0]{\Theta_{ij}^S}
\newcommand{\coupi}[0]{\Theta_{ij}^I}
\newcommand{\avg}[1]{\langle {#1} \rangle}

\newcommand{\ds}{\displaystyle}
\newcommand{\sz}{\scriptsize}
\newcommand{\nn}{\nonumber}

\newcommand{\rev}[1]{\textcolor{black}{ #1}}

\begin{document}

\title{Effective approach to epidemic containment using link equations in complex networks}

\author{Joan T.\ Matamalas}
\email{joantomas.matamalas@urv.cat}
\author{Alex Arenas}
\email{alexandre.arenas@urv.cat}
\author{Sergio G\'omez}
\email{sergio.gomez@urv.cat}
\affiliation{Departament d'Enginyeria Inform\`{a}tica i Matem\`{a}tiques, Universitat Rovira i Virgili, 43007 Tarragona, Spain}
\date{}

\begin{abstract}
Epidemic containment is a major concern when confronting large-scale infections in complex networks. Many works have been devoted to analytically understand how to restructure the network to minimize the impact of major outbreaks of infections at large scale. In many cases, the strategies consist in the isolation of certain nodes, while less attention has been paid to the intervention on links. In epidemic spreading, links inform about the probability of carrying the contagion of the disease from infected to susceptible individuals. Note that these states depend on the full structure of the network, and its determination is not straightforward from the knowledge of nodes' states. Here, we confront this challenge and propose a set of discrete-time governing equations \rev{which} can be closed and analyzed, assessing the contribution of links to spreading processes in complex networks. Our approach allows an scheme for the \rev{containment} of epidemics, based on deactivating the most important links in transmitting the disease. The model is validated in synthetic and real networks, obtaining an accurate determination of the epidemic incidence and the critical thresholds. Epidemic containment based on links' deactivation promises to be an effective tool to maintain functionality on networks while controlling the spread of diseases, as for example in air transportation networks. \\ \\

\begin{center}
  \textit{Science Advances} \textbf{4} (2018) eaau4212 \\
  \url{http://doi.org/10.1126/sciadv.aau4212}
\end{center}

\end{abstract}

\maketitle


\section*{Introduction}

The problem of modeling the spread of a disease among individuals has been studied in \rev{depth} over many years \cite{anderson1992infectious,hethcote2000mathematics,daley2001epidemic,pastor2015epidemic}. The development of compartmental models, models \rev{which} divide the individuals among a set of possible states, has given rise to a new collection of techniques that enable, for instance, the analysis of the onset of epidemics \rev{\cite{pastor2001epidemic,newman2002spread,Hufnagel2004forecast,wang2003,chakrabarti2008epidemic,gomez2010discrete,gomez2011nonperturbative,Brockmann2013hidden,cai2014effective,cai2016solving,pare2018discrete}} or the study of the impact of a vaccination campaign \rev{\cite{Earn2000simple,pastorsatorras2002inmuni,Madar2004inmu,GomezGardenes2006inmu,hebert2013global}}. Previous works heavily rely on \rev{a} mathematical approach to the study of epidemic spreading \cite{Lofgren2014math} and here we follow the same spirit.

The design of effective containment strategies constitutes a major challenge. Measures like vaccination, \rev{improved hygiene, bio-security}, cattle \rev{culling}, or education to prevent contagions, operate on the biological aspects of the disease. On the other hand, isolation or mobility restrictions act on the physical spreading channels, which may transform a local event into a pandemic. Here, we concentrate on the role of the links of the spreading network. For example, if we identify the edges which are more involved in the propagation of a disease, it is possible to design targeted countermeasures which affect just specific links instead of whole nodes, while being more effective. This can be illustrated by a hypothetical pandemic disease propagated using the air transportation network: the isolation of one airport is a dramatic measure \rev{which} is socially and politically difficult to accept and put into practice, but the suspension of just a few connections between selected airports could be more easily assumed, and at the same time achieving a better \rev{containment} of the disease.

Previous works have directed their attention mostly towards schemes based on the actuation on single nodes, either randomly or according to node properties such as their degree, betweenness, PageRank or eigenvector centrality~\cite{holme2002attack,cohen2003efficient,chung2009distributing,prakash2013fractional}. Following the same idea, some authors have introduced link removal using properties of the adjacent nodes (degrees or centralities) or of the link itself (edge betweenness)~\cite{holme2002attack,schneider2011suppressing,van2011decreasing}. A model of coevolution of epidemics with permanent and temporal link removals was proposed in~\cite{zanette2008infection}, \rev{and methods from optimization and control have been applied to minimize the impact of the epidemics~\cite{preciado2014optimal,zargham2014worst,nowzari2016analysis}}. Currently, it is considered that the optimal approach consists in finding the minimum set of edges whose removal leads to a maximum decrease of the spectral radius of the network, i.e., the largest eigenvalue of the adjacency matrix~\cite{bishop2011link,van2011decreasing,saha2015approximation}. Since the epidemic threshold is, at first order approximation of a susceptible-infected-susceptible (SIS) epidemic dynamics, inverse to the spectral radius, it seems the best and more mathematically grounded option. Unfortunately, it turns out to be an NP-complete problem, thus only heuristics are available for large networks~\cite{van2011decreasing}.

It must be emphasized that all the previous approaches make use only of the structural characteristics of the network to decide which nodes or edges have to be removed; the characteristics or parameters of the epidemic process are ignored. Even the spectral radius, which is closely related to the epidemic threshold, does not depend on the infection or recovery rates, the expected number of infected neighbors around a certain node, or any other local or global information of the spreading process.

Our proposal concentrates on the role of the links in the spreading of the epidemics, quantifying the importance of each link~\cite{restrepo2006characterizing}, thus enabling containment strategies based on their removal. To this end, we first define the epidemic importance of a link as its capacity to infect other individuals once this link has been used to propagate the disease. The determination of this link epidemic importance requires the development of a mathematical model able to cope with the infection propagation at the level of links in complex networks. We will show \rev{that the proposed model} facilitates the determination of the epidemics incidence and threshold with high accuracy. Moreover, the quantification of the \rev{epidemic} importance of the edges leads to a link removal strategy which in many cases outperforms the previous approaches, even those based on minimizing the spectral radius, and preserving at the same time most of the connectivity of the network.

The contributed model is built upon the relationships between the states of nodes connected by links, thus being related to pairwise approximations~\cite{van2009virus,taylor2011,cator2012second,kiss2012modelling,gleeson2013binary,mata2013pair,szabo2016oscillating}. However, our model is microscopic, at the level of individual links as in~\cite{mata2013pair}, thus allowing for a clear identification of maximally infectious links, but with the additional advantage of being able to easily calculate the incidence of the epidemics and the importance of the links, not just the epidemic threshold.

\section*{Results}

\begin{figure*}[!tb]
    \begin{center}
        \includegraphics[width=1.0\textwidth]{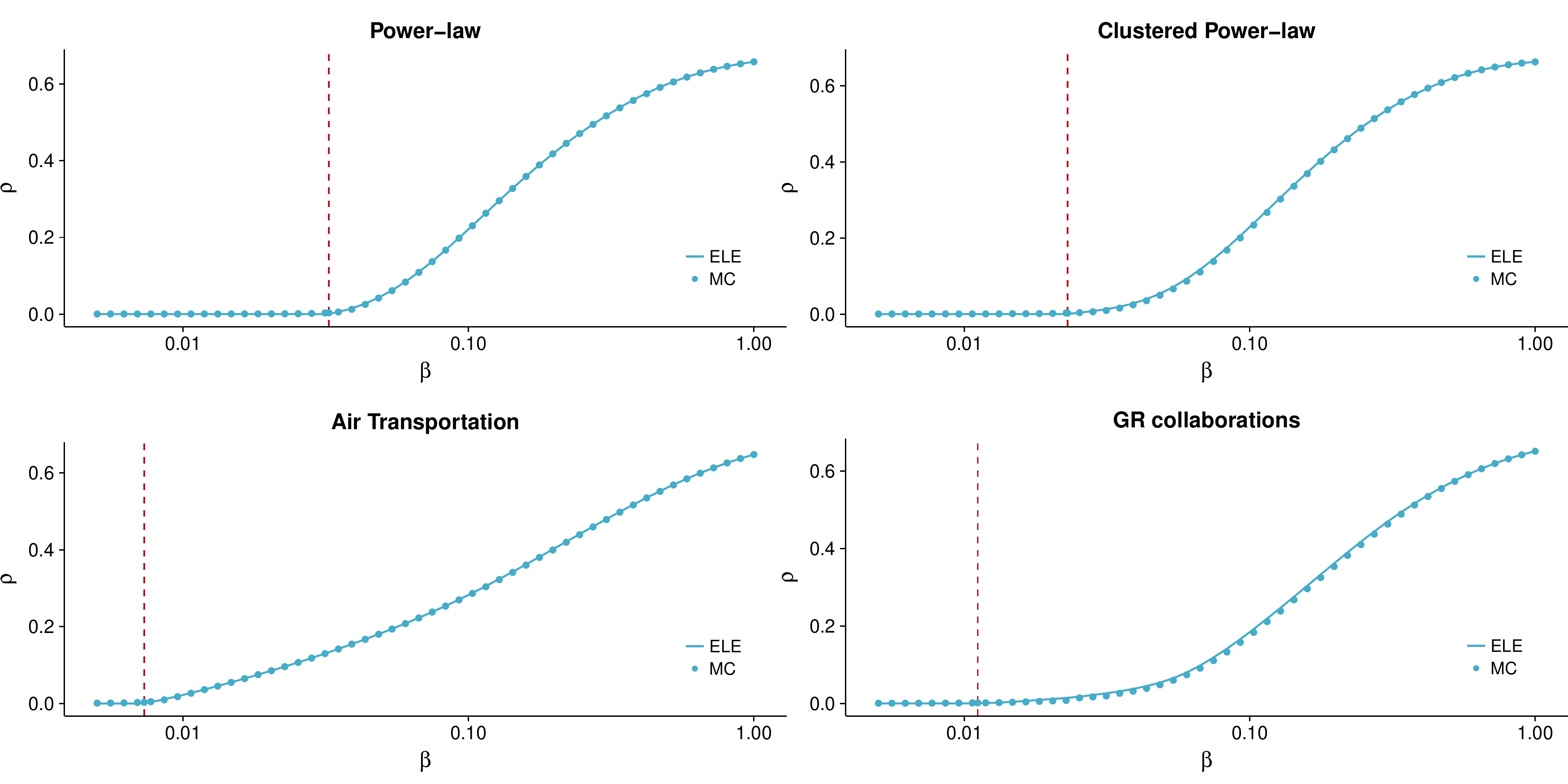}
        \caption{\textbf{Incidence of the epidemic process $\rho$ as a function of the infection probability $\beta$}. We show the incidence level for the ELE model (solid lines) and for Monte Carlo simulations (circles). The theoretical epidemic threshold calculated using equation~(\ref{eq:onset}) is \rev{marked with} a vertical line. We have made use of two synthetic and two real networks: two scale-free networks (top) with exponent 3, one of them with high clustering coefficient; the world air transportation network; and the network of scientific collaborations in the field of general relativity. We have set the recovery rate for all the networks to $\mu = 0.5$. See \textit{Methods} for the description of the networks and the details of the Monte Carlo simulations.}
        \label{fig:rho_vs_beta}
    \end{center}
\end{figure*}

\subsection*{Link Epidemic Importance}

Let us consider a discrete time SIS dynamics \rev{which} runs on top of a complex network of $N$ nodes and $L$ edges, with adjacency matrix $A$, and where each node $i$ can be in one of two different states $\sigma_i$, either susceptible ($S$) or infected ($I$), i.e.\ $\sigma_i\in\{S,I\}$. We can say that a link $(i,j)$ between nodes~$i$ and~$j$ is in state~$SI$ if $\sigma_i=S$ and $\sigma_j=I$. The parameters of the SIS dynamics are the infection and recovery probabilities, $\beta$ and $\mu$, respectively.

Our objective is to find an effective strategy to contain the SIS epidemic process through bond percolation. To determine which link should be removed first, we need a measure of the importance of each link in the spreading of the epidemics. \rev{A possible option would be to use the edge importance defined in~\cite{restrepo2006characterizing}, which accounts for the relative change of the spectral radius when the edge is removed. However, this constitutes an indirect way of containment, since we aim at lowering the incidence of the epidemics as much as possible, whereas the actuation on the spectral radius is directed to increase the epidemic threshold; both are different targets. Additionally, the spectral radius just depends on the structure of the network, but not on the epidemic parameters of the process nor the participation of the link in the spreading of the disease.}

We assume the system has reached the stationary state, which does not mean that the nodes \rev{remain in a certain fixed} state, just that the average incidence of the epidemics is \rev{basically} constant; thus, there is still \rev{margin} for applying a \rev{containment} strategy to minimize this incidence. In this regime, we can measure which are the probabilities of nodes and links to be in each of the epidemic states, e.g.\ the probability $P(\sigma_i = I)$ of node~$i$ being infected, or the joint probability $P(\sigma_i = I, \sigma_j = S)$ of link $(i,j)$ being in state~$IS$.

Consider we have a link in state $SS$ or $II$. In both cases, the next step of the epidemic dynamics is not going to use this link, since in the former there is no infected node to propagate the disease, and in the latter both nodes are already infected. Thus, to propagate the epidemics, a link must be either in an $SI$ or $IS$ state. Let us suppose we have a link $(i,j)$ in state $IS$. First, with probability $\beta$, node~$i$ can infect node~$j$ through this link, changing to state $II$. Next, infected node~$j$ may transmit the disease to some of its neighbors. Thus, if we had removed link $(i,j)$, we would have cut this path of infections initiated at node~$i$. This means that, the larger the expected number of infected neighbors of node~$j$, the largest the impact of removing link $(i,j)$ for the spreading of the epidemics. Note that not only the degree of~$j$ is relevant, but also the probability of its neighbors being susceptible when $j$ is infected, since you cannot infect nodes which are already infected. For example, if $j$ is surrounded by many infected nodes, cutting link $(i,j)$ is not going to have too much effect on the overall incidence of the epidemics. The expected number of infected nodes produced in this way can be expressed as
\begin{equation}
    \bar{n}_{ij} = \beta P(\sigma_j = S, \sigma_i = I) \sum_{r=1}^N A_{jr}\beta P(\sigma_r = S | \sigma_j = I)\,,
\end{equation}
where $P(\sigma_r = S | \sigma_j = I)$ is the conditional probability that node~$r$ is susceptible when its neighbor~$j$ is infected. Since this measure is asymmetric, and removing an edge affects the propagation of the disease in both directions, we define the \textit{link epidemic importance} of a link, $I_{ij}$, as
\begin{equation}
    \label{eq:importance}
    I_{ij} = \bar{n}_{ij} + \bar{n}_{ji}\,.
\end{equation}

Now, the problem reduces to finding the joint and conditional probabilities for each link, and this is accomplished using our \textit{Epidemic Link Equations} (ELE). It can be shown that this definition of link epidemic importance has the property of trying to preserve the connectivity of the network (see section~S1 and fig.~S1), unlike other options such as edge betweenness, which quickly tend to produce a large number of disconnected components, thus hindering the functionality of the network.

\subsection*{Epidemic Link Equations}

\begin{figure*}[!tb]
    \begin{center}
        \centering
        \includegraphics[width=1.0\textwidth]{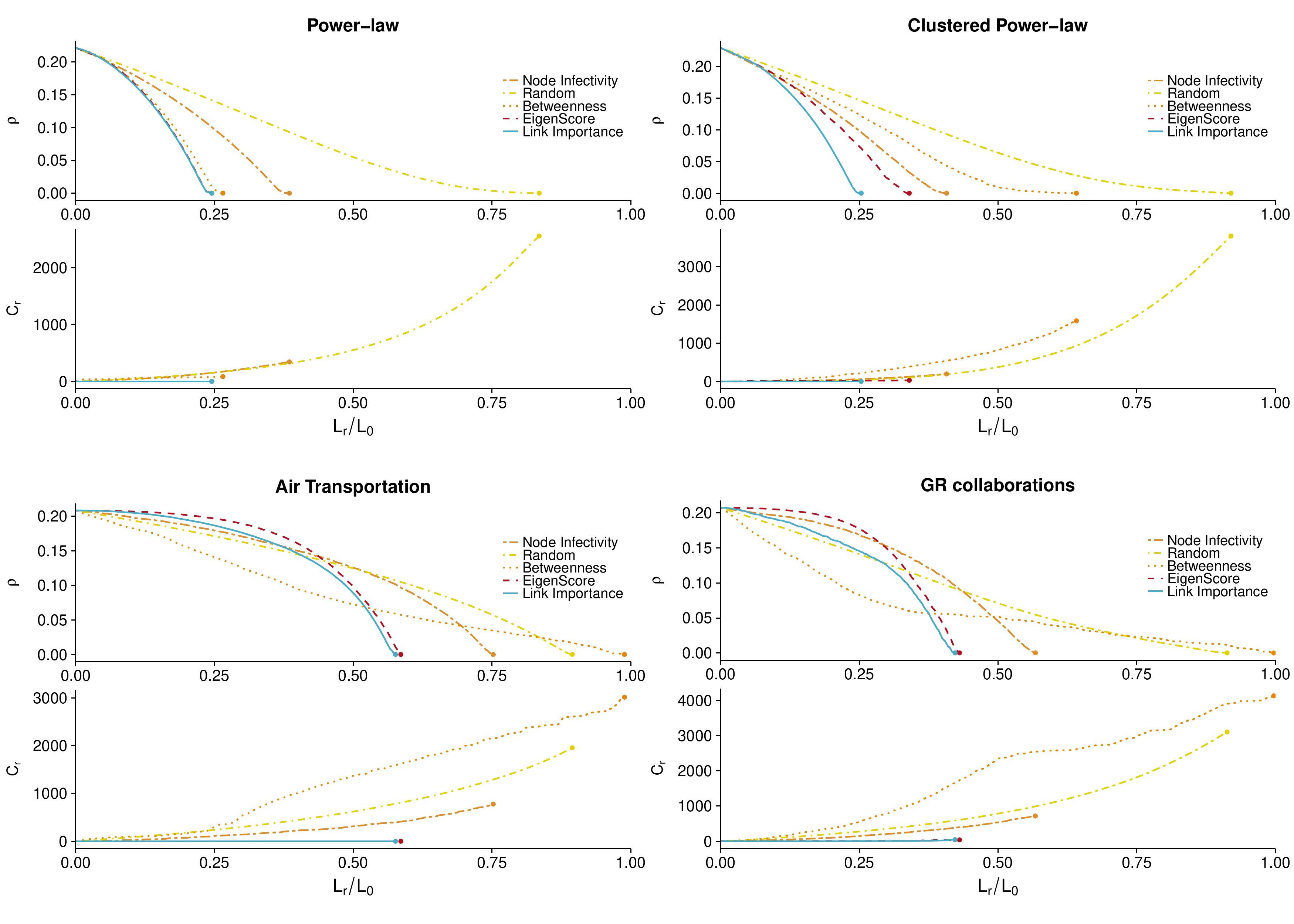}
        \caption{\textbf{Targeted bond percolation}. We show the incidence of the epidemics, $\rho$, and the number of connected components, $C_r$, as function of the occupation probability, $L_r/L_0$, where $L_0$ is the number of links of the network and $L_r$ the number of removed edges in the bond percolation process. We compare five different epidemic containment strategies: removing the edges of the node with highest probability of being infected, $P(\sigma_i = I)$ (orange dash-dash line); a random edge removal (yellow dash-dot line); removing the edge with highest edge-betweenness (light orange dotted line); targeting the edge with highest eigenscore (red dashed line); and finally, removing the edge \rev{which} has the largest \rev{link epidemic} importance (blue solid line). We apply these processes to the same networks as in Fig.~\ref{fig:rho_vs_beta} (see \textit{Methods}). We have set the recovery rate to $\mu = 0.5$ and chosen the infection probability $\beta$ such that the stationary incidence of the epidemics is about $\rho_{\mbox{\scriptsize ini}} \approx 0.2$ for all the networks, i.e.\ $\beta = 0.1$ for both power-law networks, $\beta = 0.06$ for air transportation network, and $\beta = 0.11$ for the collaboration network. The dots mark the achievement of total containment.}
        \label{fig:percolation}
    \end{center}
\end{figure*}

To simplify the notation, we first denote the previous joint probability as $\Phi_{ij}=P(\sigma_i=S,\sigma_j=I)$; the higher the $\Phi_{ij}$, the larger the likelihood that the disease propagates from node $j$ to node $i$. It is worth mentioning that this feature is in general asymmetrical, meaning that the propagation of the illness can be more probable from $j$ to $i$ than the other way around. In the same way, the epidemic is restrained by edges where the nodes are in the same state, thus it is convenient to define the probabilities $\Theta^S_{ij}=P(\sigma_i=\sigma_j=S)$ and $\Theta^I_{ij}=P(\sigma_i=\sigma_j=I)$ for all pairs of neighboring nodes.

The evolution of the joint probability $\Phi_{ij}$ of one link depends on \rev{the probabilities} $\Phi$, $\Theta^I$ and $\Theta^S$ to the rest of the neighboring links, and the infection rules of the SIS dynamics. Thus, we can write the following equation for each link:
\begin{eqnarray}
  \label{eq:Phi_mpa}
  \Phi_{ij}(t+1) & = & \Theta^S_{ij}(t)\, q_{ij}(t)\, (1 - q_{ji}(t)) \nonumber \\
                 &   & \mbox{} + \Phi_{ij}(t)\, ((1 - \beta) q_{ij}(t))\, (1 - \mu) \nonumber \\
                 &   & \mbox{} + \Phi_{ji}(t)\, \mu\, (1 - (1 - \beta)q_{ji}(t)) \nonumber \\
                 &   & \mbox{} + \Theta^I_{ij}(t)\, \mu (1 - \mu)
\end{eqnarray}
\rev{where $q_{ij}(t)$ stands for the probability that a susceptible node $i$ is not infected by any of its neighbors (excluding node $j$)}. We have taken into account all the possible changes of state of the nodes $i$ and $j$. The first term considers the probability that both nodes are in a susceptible state, and then node $i$ remaining susceptible while node $j$ is infected by any of its other neighbors. The second term accounts for both nodes remaining in the same state, node $i$ is not infected by any of its neighbors and node $j$ is not recovered from the infection. Then, the third term represents the transition in which node $i$ is infected and recovers while node $j$ is susceptible and it is infected by any of its other neighbors. Finally, in the fourth term both nodes are infected but node $i$ recovers while node $j$ does not. The asymmetry of probability $\Phi_{ij}$ multiplies the number of equations by two, since for each link between nodes $i$ and $j$ we need an equation for $\Phi_{ij}(t+1)$ and another for $\Phi_{ji}(t+1)$.

Similarly we can obtain an expression for probability $\Theta_{ij}^I$:
\begin{eqnarray}
  \label{eq:Theta_mpa}
  \Theta_{ij}^I(t+1) & = & \Theta^S_{ij}(t)\, (1-q_{ij}(t))\, (1 - q_{ji}(t)) \nonumber \\
                     &   & \mbox{} + \Phi_{ij}(t)\, (1 - (1 - \beta)q_{ij}(t))\, (1 - \mu) \nonumber \\
                     &   & \mbox{} + \Phi_{ji}(t)\, (1 - \mu)\, (1 - (1 - \beta)q_{ji}(t)) \nonumber \\
                     &   & \mbox{} + \Theta^I_{ij}(t)\, (1 - \mu)^2
\end{eqnarray}
In this case we have only $L$ equations, one per link, due to its symmetry. There is no need of extra equations for probability $\Theta^S_{ij}$ since the normalization leads to $\Theta^S_{ij} = 1 -\Phi_{ij}-\Phi_{ji}-\Theta^I_{ij}$.

The $q_{ij}(t)$ in equations~(\ref{eq:Phi_mpa}) and~(\ref{eq:Theta_mpa}) \rev{can be expressed as}:
\begin{equation}
    q_{ij}(t)  = \prod_{\substack{r = 1\\r\neq j}}^{N} \left( 1 - \beta A_{ri} h_{ir}\right)
\end{equation}
where $h_{ij}$ defines the hostility of $j$ against $i$, i.e.\ the probability that node $j$ is infected when node $i$ is susceptible, $h_{ij}=P(\sigma_j=I | \sigma_i =S)$. The hostility can be obtained in terms of $\Theta_{ij}^S$ and $\Phi_{ij}$ as:
\begin{equation}
    \label{eq:hostility}
    h_{ij} = \frac{\Phi_{ij}}{\Phi_{ij} + \Theta_{ij}^S}
\end{equation}
Note that the denominator in equation~(\ref{eq:hostility}) is a property of node $i$ given that $\Phi_{ij} + \Theta_{ij}^S = P(\sigma_i = S)$ for all neighboring nodes $j$ of vertex $i$.

We call this system of $3L$ equations and unknowns our \textit{Epidemic Link Equations} (ELE) model. It can be solved by iteration, starting from any meaningful initial condition, e.g.\ $\coupi(0)=\rho_0^2$ and $\condij(0)=\condji(0)=\rho_0(1-\rho_0)$ (for any $0<\rho_0\le 1$), until \rev{a fixed point} is found. Apart from the solution where all nodes are susceptible, $\coups = 1$ for all the links, a non-trivial one appears when the system is above the critical value of the epidemic spreading (see \textit{Methods} for the analytic derivation of the epidemic threshold from ELE model). Finally, the incidence of the epidemic process, the average number of infected nodes in the whole system, can be computed as:
\begin{equation}
    \rho = \frac{1}{N}\sum_{i=1}^N \frac{1}{k_i} \sum_{j=1}^{N} A_{ji}(\condji + \coupi)
\end{equation}
where $k_i$ is the degree of node $i$.

To test the agreement between our approach and empirical simulations we have analyzed the incidence of the epidemics, $\rho$, in different synthetic and real network structures, covering the full range of infection probabilities, $\beta$, see Fig.~\ref{fig:rho_vs_beta}. The results show a remarkable agreement between our ELE model and the Monte Carlo simulations, and a good prediction of the epidemic threshold for all synthetic and real networks, pointing out the validity of our model to describe the global impact of the epidemics. Note that \rev{all networks}, except the first one, have a large clustering coefficient, making the determination of the incidence difficult for standard mean field methods, due to the effect of dynamical correlations.

\begin{figure*}[!tb]
    \begin{center}
        \centering
        \includegraphics[width=1.0\textwidth]{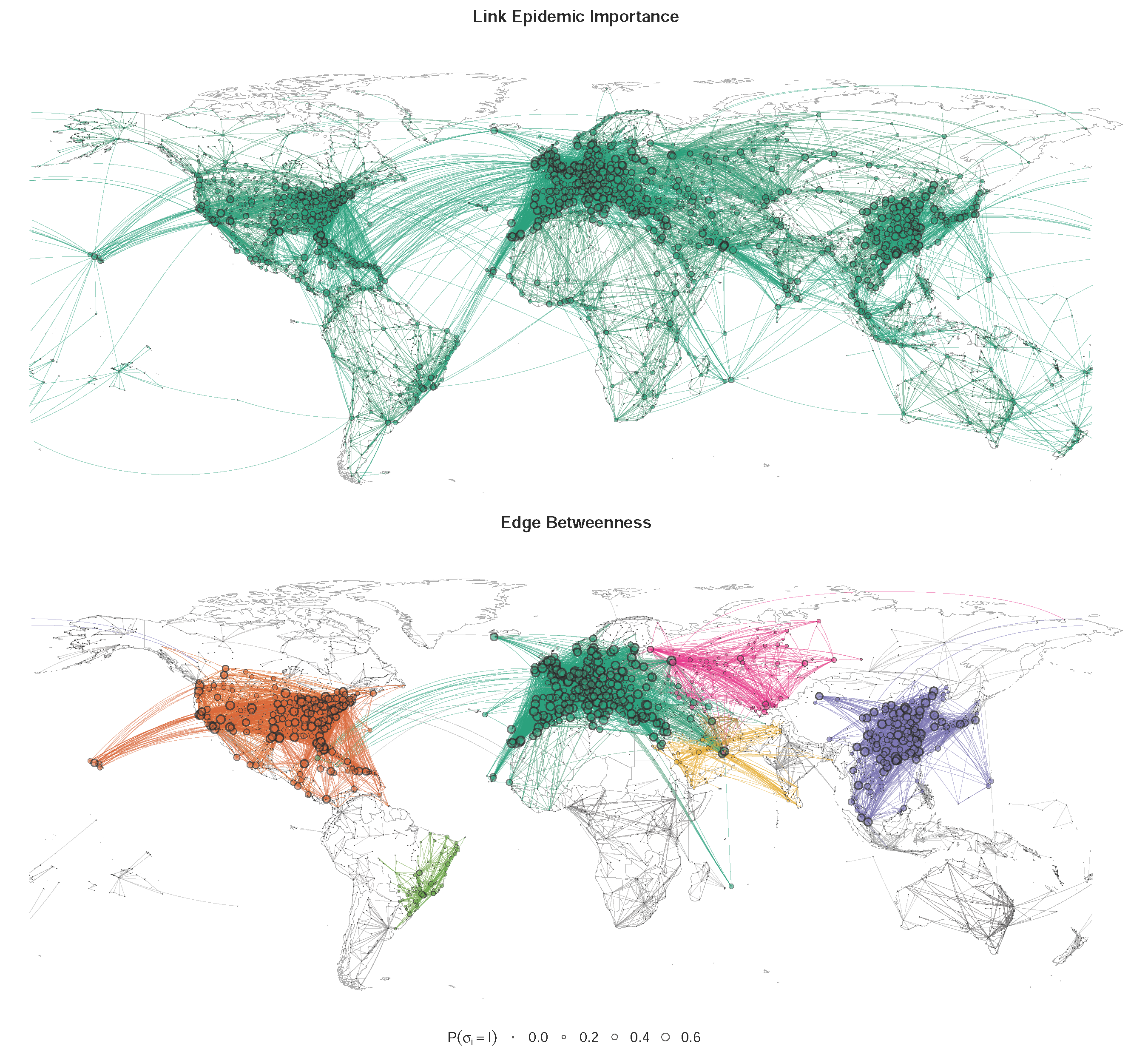}
        \caption{\textbf{Epidemic containment on the air transportation network}. We show the networks after 33.3\% of the links have been removed using link epidemic importance (top) and edge betweenness (bottom). Nodes and edges with the same color belong to the same connected component, with subcritical components in gray scale and using darker gray for larger components. The area of the nodes is proportional to their probability of being infected $P(\sigma_i=I)$ from $0.0$ to $0.6$. We have set the epidemic probabilities to $\mu = 0.5$ and $\beta = 0.06$.}
        \label{fig:world}
    \end{center}
\end{figure*}

\begin{figure*}[!tb]
    \begin{center}
        \centering
        \includegraphics[width=1.0\textwidth]{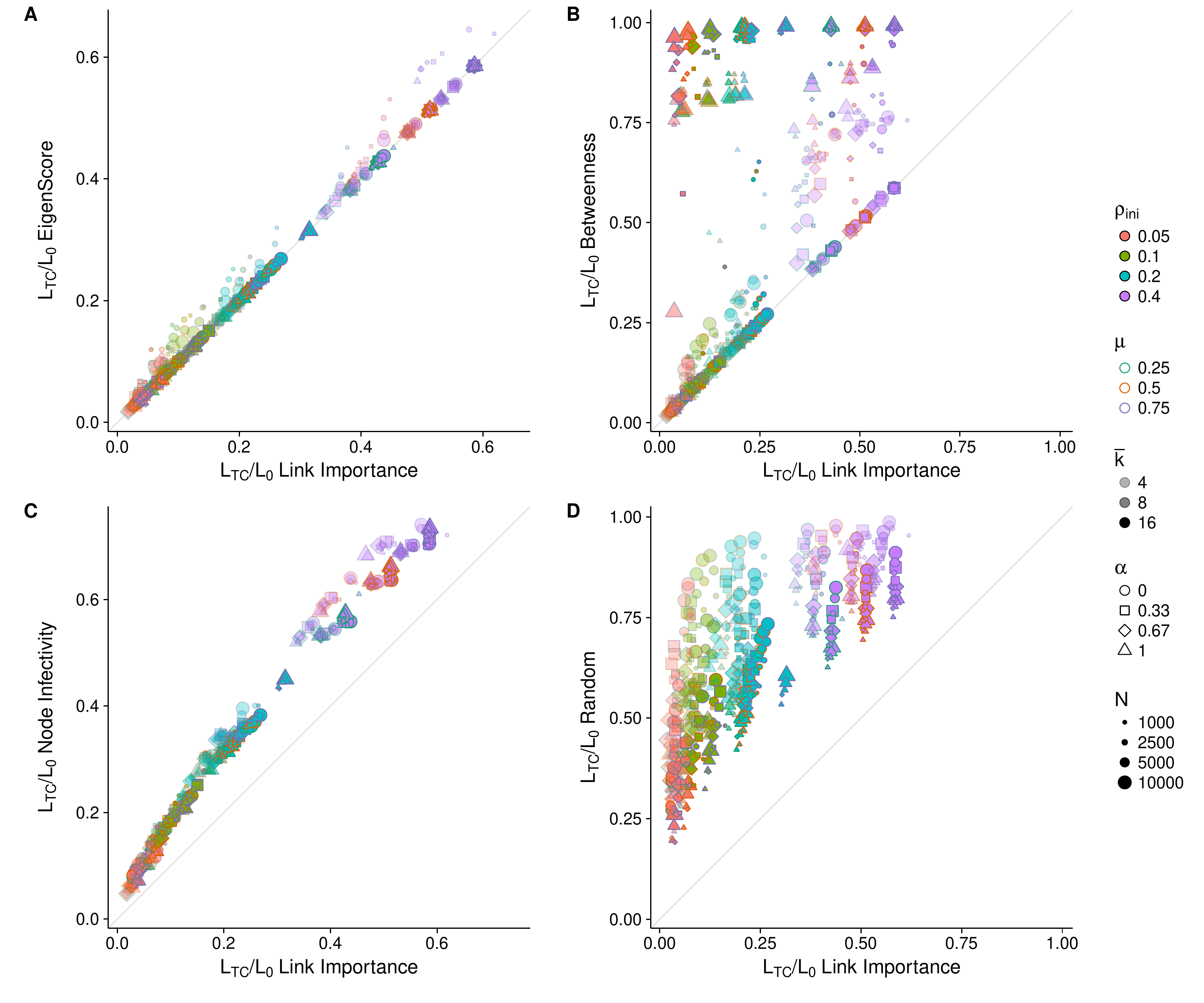}
        \caption{\textbf{Fraction of links removed for total epidemic containment on synthetic networks}. We show the fraction of links \rev{which} have to be removed to obtain total epidemic containment using \rev{link} epidemic importance, compared with the fractions for the other four strategies: (A) eigenscore; (B) edge betweenness; (C) node infectivity; (D) random removal. Each point represents a configuration consisting of a network and a set of epidemic parameters. The networks have been generated with the model in \cite{gomez2006scale}, which interpolates between Erd\H{o}s-R\'enyi ($\alpha=1$) and Barab\'asi-Albert networks ($\alpha=0$). We use four values of the interpolating parameter: $\alpha=0.0$, $0.33$, $0.67$, $1.0$. These networks are generated in four sizes ($N=1000$, $2500$, $5000$, $10000$) and three average degrees ($\avg{k}=4$, $8$, $16$), thus amounting 48~different networks. For each network, we apply the five containment strategies for three different values of the recovery probability ($\mu=0.25$, $0.50$, $0.75$), and four values of the infection probability $\beta$ selected such that, before removing links, the incidence of the epidemics at the stationary state is equal to $\rho_{\mbox{\scriptsize ini}}=0.05$, $0.1$, $0.2$, $0.4$. Therefore, each plot contains 576~different configurations.}
        \label{fig:synthetic}
    \end{center}
\end{figure*}

\subsection*{Epidemic Containment}

Our approach for effective epidemic containment consists in removing the links with largest \textit{link epidemic importance}. This is possible once we have solved the ELE model, computing the $I_{ij}$ for all the links in the network using equation~(\ref{eq:importance}), which can be expressed as
\begin{equation}
    \label{eq:lei}
    I_{ij} = \beta^2 \left( \Phi_{ji} \sum_{r=1}^N A_{jr} \frac{\Phi_{rj}}{\Phi_{rj} + \Theta^I_{rj}}
                          + \Phi_{ij} \sum_{r=1}^N A_{ir} \frac{\Phi_{ri}}{\Phi_{ri} + \Theta^I_{ri}}
                     \right)\,.
\end{equation}
Note that the value of $\beta$ does not affect the ranking of the links, but we \rev{do not remove it from equation~(\ref{eq:lei}) to preserve} the semantics of $I_{ij}$.  Since the structure of the network changes after each link removal, it is convenient to recalculate the solution of the ELE model to ensure that we really remove the current link with largest \rev{link epidemic} importance.

\begin{figure}[!tb]
    \begin{center}
        \centering
        \includegraphics[width=0.8\columnwidth]{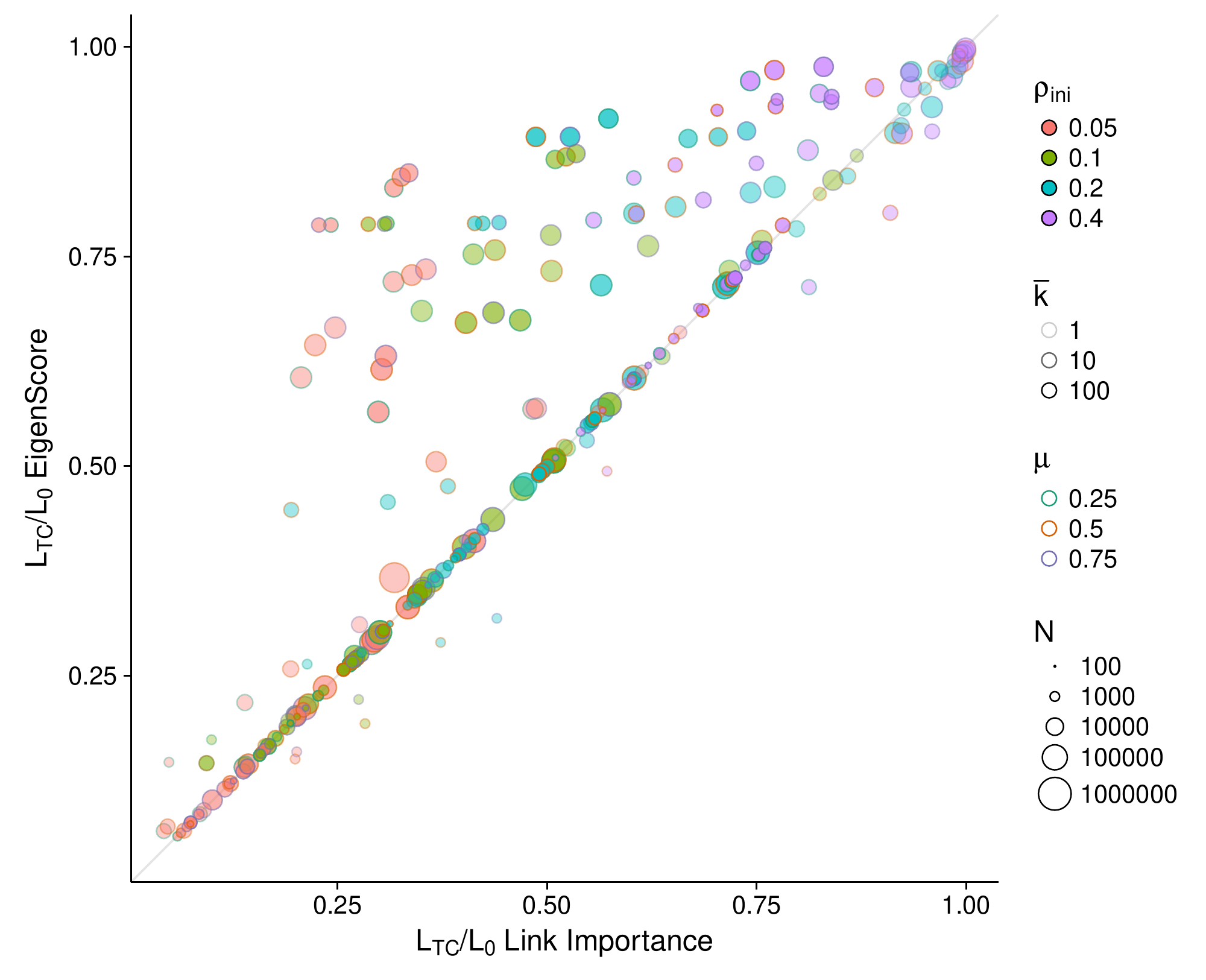}
        \caption{\textbf{Fraction of links removed for total epidemic containment on real networks}. We compare the \rev{link} epidemic importance and eigenscore methods on a set of 27~real networks selected from the Network Repository \cite{rossi2015network} (see \textit{Methods}), with sizes ranging from 410~to 404719~nodes. The epidemic parameters are the same as in Fig.~\ref{fig:synthetic}, thus amounting 324~different configurations.}
        \label{fig:real}
    \end{center}
\end{figure}

We show the results of our approach for epidemic containment in Fig.~\ref{fig:percolation}. For comparison purposes, we also test four additional containment strategies. First, two strategies \rev{which} just consider the structure of the network: removal based on maximal edge betweenness~\cite{holme2002attack}, and targeting the link with highest eigenscore, i.e., the product of the eigenvector centralities of the nodes connected by the link~\cite{van2011decreasing}. Then, we consider a measure based on the epidemic process at the level of nodes, the removal of all the links of the node \rev{which} has maximal probability of being infected. Finally, a simple random edge removal. As in the case of our strategy, we recalculate all the measures after each removal (see \textit{Methods} for further details). \rev{We have also checked a promising approach based on communicability distances \cite{estrada2015}, however the computational costs involved in computing communicability angles and distances for large networks preclude this approach to be used in large networks, for this reason we have not included results on this one.}

We observe in Fig.~\ref{fig:percolation} that link \rev{epidemic} importance leads to the fastest extinction of the epidemics for the four considered networks, and it is the only method \rev{which} preserves their connectivity (thus, functionality). Note that the strategy based on node infectivity performs poorly for all the networks despite having information about the epidemic process, although better than the random removal. This means that the use of information at the level of links is crucial to contain the epidemics.

For the power-law network, our approach using link \rev{epidemic} importance \rev{yields} the best performance, but the results are very similar to the ones obtained using eigenscore and edge betweenness strategies (equivalent results hold for Erd\H{o}s-R\'enyi networks, see fig.~S6). However, when the transitivity of the network is increased, we can clearly see the benefits of using link \rev{epidemic} importance, both in epidemic containment and on preservation of the connectivity of the network (see figs.~S2 and~S3 for more details on the containment process for each method).

The effect of the clustering coefficient is also present when we look \rev{at} the epidemic containment results for the two empirical networks in Fig.~\ref{fig:percolation}. Moreover, as in most real networks, the air transportation and the scientific collaborations networks have a significant modular structure. This plays an important role on the epidemic containment process. Here, we can see how the strategy based on edge betweenness apparently performs better when few links are removed, due to the fact that links with higher edge betweenness are those connecting different modules~\cite{newman2004finding}. When the bond percolation process isolates modules, each module may sustain its own epidemic process, and thus it may happen that some of the modules are subcritical for the given infection probability $\beta$. That will lead to a decrease of the global prevalence of the epidemics at expenses of loosing the connectivity of the network. Furthermore, if we look at the prevalence on the giant connected component, an important increase above the initial average number of infected individuals is revealed (see figs.~S4 and~S5). A consequence of this fragmentation process is the appearance of multiple isolated supercritical components, for which the removal of a link in one of them does not affect the incidence on the other components. As a result, the edge betweenness procedure needs to remove more links to arrive to the total epidemic extinction than any of the other methods, even the random one. For the sake of completeness we have analyzed two benchmark networks with community structure, obtaining similar results (see figs.~S7 and~S8).

In Fig.~\ref{fig:world} (top) we illustrate the survival links in the air transportation network after 33.3\% of the edges have been removed according to our epidemic containment strategy proposal (see fig.~S9 for the original network before the containment process). As it is observed, the global connectivity, and thus functionality, of the worldwide connections is preserved (links of the same color are part of the same connected component). In Fig.~\ref{fig:world} (bottom) we plot the network after deactivating the same fraction of links (33.3\%) using the recursive deactivation of links according to edge betweenness. The edge betweenness containment method, in contrast with our proposal, generates two main kinds of components: small or sparsely connected subcritical modules like the ones in Australia, Africa or South America, where the epidemics vanishes; and large supercritical communities in Europe, North America and Easter Asia, with a large prevalence of the epidemics. This means that, for instance, there is no path to go from London to New York, or from Tokyo to Los Angeles, thus disconnecting the world by air transportation.

For a better assessment of the performance of the different containment strategies, we show in Fig.~\ref{fig:synthetic} their comparison in terms of the required fraction of removed links to attain total containment, $L_{TC}/L_0$, when applied to a large set of synthetic networks and epidemic parameters. The results point to a clear advantage of the link \rev{epidemic} importance method over the node infectivity and random approaches, and better or equal results with respect to edge betweenness and eigenscore. In fact, only eigenscore achieves results comparable to link \rev{epidemic} importance, with a slight advantage for our method. When applied to a set of 27~real networks (see \textit{Methods}), the differences between \rev{link} epidemic importance and eigenscore become more evident, showing again the effectiveness of our approach, but with some exceptions in which eigenscore performs better, see Fig.~\ref{fig:real}. Additionally, we plot in figs.~S10 and~S11 the comparison of the number of connected components between the different containment strategies, which show the better \rev{performance} of link \rev{epidemic} importance to keep low the number of components, with just a few exceptions.

\section*{Discussion}

We have presented a methodology for assessing epidemic spreading based on links instead of nodes. The model, named Epidemic Link Equations (ELE), allows the determination of the \rev{epidemic} importance of each link in transmitting the disease. The method accounts for the first order correlations between links, although it could be extended to higher orders assuming a larger analytical and computational cost. The results are used to develop an epidemic containment strategy consisting in deactivating recursively the \rev{links with} largest \rev{link} epidemic importance while preserving the connectivity of the full network, i.e. avoiding fragmentation. We have validated our proposal in synthetic and empirical networks, comparing with other alternative containment strategies, \rev{which show its better performance, with few exceptions}. In the empirical case of the worldwide air transportation network, we identify the most important connections between airports for the spreading of epidemics and evaluate the epidemic incidence after its deactivation, considering an SIS epidemic spreading dynamics. Our results open the door to new approaches in the analysis of dynamical diffusive-like models on complex networks at the level of links instead of nodes.

\section*{Methods}

\subsection*{Epidemic Threshold}

The determination of the epidemic threshold is performed by considering a state of the system in which the epidemic incidence is very small ($\condij,\condji,\coupi\ll 1$, for all links), thus the system of equations can be linearized (See section~S2 for full details), resulting in:
\begin{eqnarray}
    \label{eq:coupi_lin}
    \coupi &=&\beta(1-\mu)\condij + \beta(1-\mu)\condji + (1-\mu)^2\coupi
    \\
    \condij &=&\beta\sum_r(A_{rj}-(1-\mu)\delta_{ri})\condsub{jr}
    \nonumber
    \\
    \label{eq:condij_lin}
            & & \mbox{} + (1-\beta)(1-\mu)\condij + \mu(1-\mu)\coupi
\end{eqnarray}
Here we have removed the dependence on time, to emphasize we are considering the steady state. From equation~(\ref{eq:coupi_lin}) we can write
\begin{equation}
    \label{eq:coupi_sum_conductance}
    \coupi = \frac{\beta(1-\mu)}{\mu(2-\mu)}(\condij+\condji)
\end{equation}
Now, calling $\varepsilon_i=\condji+\coupi\ll 1$, which does not depend on node $j$ since $P(\sigma_i=I,\sigma_j=S)+P(\sigma_i=I,\sigma_j=I)=P(\sigma_i=I)$, we make the following ansatz:
\begin{eqnarray}
    \coupi & = & \Upsilon(\varepsilon_i+\varepsilon_j)
    \\
    \condij & = & X\varepsilon_i+Z\varepsilon_j
\end{eqnarray}
where $\Upsilon$, $X$ and $Z$ are constants independent of the link. These ansatz include the assumption of symmetry $\coupi$ and asymmetry of $\condij$, respectively. We can determine the constants by substitution in equation~(\ref{eq:coupi_sum_conductance}) and using the definition of $\varepsilon_i$, which leads to
\begin{eqnarray}
    \Upsilon & = & \frac{\beta(1-\mu)}{\mu(2-\mu)+2\beta(1-\mu)}
    \\
    X & = & -\Upsilon
    \\
    Z & = & 1 - \Upsilon
\end{eqnarray}
Finally, we build equations for the $\varepsilon_i$ by substituting equations~(\ref{eq:coupi_lin}) and~(\ref{eq:condij_lin}) in $\varepsilon_i=\condji+\coupi$, and using the ansatz. The result is
\begin{equation}
    \label{eq:theta_value}
    \frac{\mu}{\beta}\varepsilon_i = \sum_j B_{ji}\varepsilon_j
\end{equation}
where $B$ is a matrix whose elements depend on the adjacency matrix of the network, on $\Upsilon$ and on the degrees $k_i$ of the nodes:
\begin{equation}
    \label{eq:matrix_b}
    B_{ij} = (1 - \Upsilon)A_{ij} - \Upsilon k_i\delta_{ij}
\end{equation}
The $\delta_{ij}$ stands for the Kronecker delta function, which is 1 if $i=j$, and 0 otherwise. If $\mu/\beta$ is an eigenvalue of matrix $B$, equation~(\ref{eq:theta_value}) has a non-trivial solutions. Hence, the onset of the epidemics $\beta_c$, the lowest value of $\beta$ \rev{which} yields non-trivial solutions of equation~(\ref{eq:theta_value}), is given by
\begin{equation}
    \label{eq:onset}
    \beta_c = \frac{\mu}{\Lambda_{\max}(B)}
\end{equation}
where $\Lambda_{\max}(B)$ is the largest eigenvalue of matrix $B$. Note that matrix $B$ depends on $\beta$ and $\mu$, thus equation~(\ref{eq:onset}) is implicit for $\beta_c$, which can be solved by iteration. See section~S3 for a discussion on the determination of the epidemic threshold.

\subsection*{Estimation of the incidence of the epidemic from numerical simulations}

The numerical incidence of the epidemics, $\rho$, is calculated using discrete-time and synchronous Monte Carlo simulations. We make use of the quasistationary approach (QS)~\cite{ferreira2012epidemic, binder1993monte} to avoid the effect that large number of realizations end up in the absorbing state with no infected individuals in the system. Basically, the QS method focus the simulation on active configurations, i.e.\ with one or more infected individuals. Every time the system reaches the absorbing state, this state is replaced by one of the previously stored active states of the system. We keep 50~active configurations with an update probability of~$0.20$. We give the systems a transient time of~$10^5$ time steps and then, we calculate the $\rho$ as an average over a relaxation time of~$2 \times 10^4$ time steps.

\subsection*{Networks}
In this work, we evaluate our methodology on synthetic and empirical networks. We have built a network with power-law degree distribution $P(k)\sim k^{-\gamma}$ with exponent $\gamma=3$ and $\langle k \rangle=6$ using the configuration model. To evaluate the impact of transitivity we have also built another network with the same characteristics of the previous one but with a clustering coefficient of $0.6$ using the algorithm by Holme et al.~\cite{holme2002growing} with a parameter $p = 0.8$.

We consider also two empirical networks: the air transportation network, and the network of scientific collaborations in the field of general relativity. The air transportation network has been constructed using data from website openflights.org which has information about the traffic between airports updated to~2012. This network accounts for the largest connected component, with 3154~nodes and 18592~edges (see data file~S1). The network of scientific collaborations has been obtained from~\cite{leskovec2007graph}; it is composed by 5242 nodes linked by 14496 edges.

The synthetic networks in Fig.~\ref{fig:synthetic} have been generated with the model in \cite{gomez2006scale}, which interpolates between Erd\H{o}s-R\'enyi (ER) and Barab\'asi-Albert (BA) networks. In this way, we are able to evaluate the performance of the containment strategies on networks with degree distributions which range from Poisson (ER) to power-law (BA). By constructon, these networks have no community structure and low transitivity.

The 27~real networks in Fig.~\ref{fig:real} have been obtained from the Network Repository \cite{rossi2015network}, selecting only the largest connected component. They cover wide ranges of number of nodes (from 410~to 404719), number of links (from 1043~to 713319), average degree (between~2.04 and~84.82), average clustering coefficient (from~0.0023 to~0.1105) and assortativity (between~-0.88 and~0.64), see section~S4.

\subsection*{Containment process}
To perform the deactivation of links, we impose an adiabatic process: after each removal step we let the system converge to the meta-stable equilibrium before removing any other link. For a fair comparison between different containment strategies, we remove on each deactivation step as many edges as we have removed using the node infectivity strategy. In the case of the real networks in Fig.~\ref{fig:real}, we have set the maximum number of adiabatic processes to~1000, due to its large computational cost on the largest networks. This means that, if the network has 20000~links, we remove 20~links at each deactivation steps. We consider we have reached total containment when the incidence of the epidemics becomes lower than $1/N$. The computational cost of calculating the links to remove for each containment strategy is shown in fig.~S13.

\section*{Acknowledgments}
We acknowledge comments on the manuscript by C.\ Granell, M.\ Barahona and P.\ H\"{o}vel. JTM, AA and SG have been supported by the Generalitat de Catalunya project 2017-SGR-896, Spanish MINECO project FIS2015-71582-C2-1, and Universitat Rovira i Virgili project 2017PFR-URV-B2-41. AA acknowledges financial support from the ICREA Academia and the James S.\ McDonnell Foundation. JTM acknowledges a FI-2015 grant from Generalitat de Catalunya AGAUR.

\section*{Contributions}
All authors have contributed equally to this manuscript.

\section*{Competing interests}
All authors declare that they have no competing interests.


\clearpage

\section*{Supplementary materials}

\renewcommand*{\thefigure}{{\bf S\arabic{figure}}}
\renewcommand{\figurename}{{\bf Fig.}}
\renewcommand*{\thetable}{{\bf S\arabic{table}}}
\renewcommand{\tablename}{{\bf Table}}
\renewcommand{\refname}{Supplementary References}
\renewcommand{\theequation}{S.\arabic{equation}}

\renewcommand{\thesection}{S\arabic{section}}
\titleformat{\section}[block]
    {\bfseries}
    {Section \thesection.}
    {2ex}{}

\setcounter{equation}{0}
\setcounter{figure}{0}

\section{Link \rev{epidemic} importance and connected components} \label{sec:link_importance}

One important property of our proposed definition of \textit{link epidemic importance} is that it tends to maintain the connectivity of the network when the selected link is removed. Here we show that, when the network is formed by two subnetworks $A$ and $B$, which are connected by just one link $(i_A,j_B)$, the \rev{link epidemic} importance of that link is lower than \rev{that} of another link internal to $A$ or $B$. This means our containment strategy of removing the edge with largest link epidemic importance will not break the network in two disconnected components $A$ and $B$, unlike the betweenness approach, for which the link $(i_A,j_B)$ plays the role of a bridge and thus it has maximal edge betweenness.

Let $\avg{k}_A$ and $\avg{k}_B$ be the average degrees of subnetworks $A$ and $B$, respectively, with $\avg{k}_A > \avg{k}_B$. Let us also call $\rho^A$ and $\rho^B$ their respective incidence of the epidemics. The link epidemic importance of a link between nodes $i$ and $j$ has been defined as
\begin{equation}
   \label{seq:importance}
   I_{ij} = \bar{n}_{ij} + \bar{n}_{ji}\,,
\end{equation}
where
\begin{equation}
   \label{seq:importance_dir}
   \bar{n}_{ij} = \beta P(\sigma_j = S, \sigma_i = I) \sum_{r=1}^N A_{jr}\beta P(\sigma_r = S | \sigma_j = I)\,.
\end{equation}
Supposing independence of the states of the nodes, $\bar{n}_{ij}$ can be approximated as
\begin{equation}
   \label{seq:importance_aprox}
   \bar{n}_{ij} \approx \beta P(\sigma_j = S) P(\sigma_i = I) \sum_{r=1}^N A_{jr}\beta P(\sigma_r = S)\,.
\end{equation}
In a homogeneous mean field approximation we may substitute $k_j = \avg{k}$, $P(\sigma_i = I) \approx \rho$, and $P(\sigma_j = S) \approx 1-\rho$, which lead to the following expressions for the importance of a link:
\begin{eqnarray}
   \label{seq:importanceA}
   I^{A} & \approx & 2 \beta^2 \rho^A (1-\rho^A)^2 \avg{k}_A\,,
   \\
   \label{seq:importanceB}
   I^{B} & \approx & 2 \beta^2 \rho^B (1-\rho^B)^2 \avg{k}_B\,,
   \\
   \label{seq:importanceAB}
   I^{AB} & \approx & \beta^2 \left[ \rho^A (1-\rho^B)^2 \avg{k}_B + \rho^B (1-\rho^A)^2 \avg{k}_A \right]\,.
\end{eqnarray}
Here, $I^{A}$ and $I^{B}$ denote the \rev{link} epidemic importance of links inside $A$ and $B$, respectively, and $I^{AB}$ the \rev{link epidemic} importance of the link connecting subnetworks $A$ and $B$.

\begin{figure}[!t]
\centering
   \includegraphics[width=15cm]{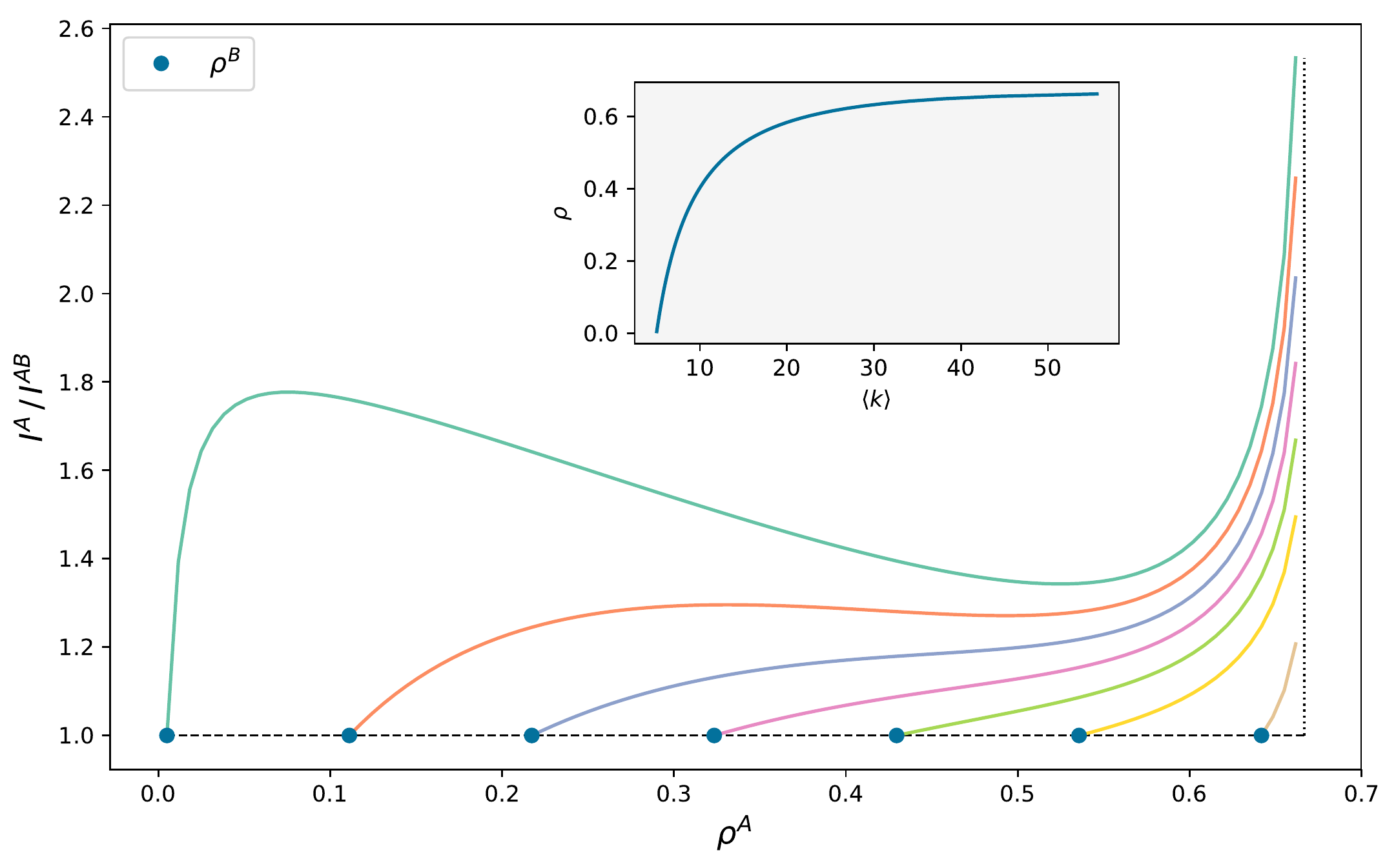}
   \caption{Ratio between the \rev{link epidemic} importance $I^A$ of a link in a subnetwork $A$ and the \rev{link epidemic} importance $I^{AB}$ of a link that acts as the only bridge between subnetworks $A$ and $B$. First, we fix the average degree $\avg{k}_B$ of subnetwork $B$ (or equivalently, we fix its incidence $\rho^B$, the red circles), and then we consider subnetworks $A$ with average degree (and epidemic incidence) larger than that of $B$, i.e.\ $\avg{k}_A \geqslant \avg{k}_B$ (thus, $\rho^A \geqslant \rho^B$). We can see that, in all cases, $I^{A} \geqslant I^{AB}$, meaning that the ranking by \rev{link} epidemic importance will not be leaded by the bridges. The vertical dotted line highlights the asymptote at $\rho=1/(1+\mu)$. The inset shows the relationship between the incidence and the average degree. We have set the epidemic parameters to $\mu=0.5$ and $\beta=0.1$, and the calculations rely on a nonperturbative homogeneous mean field approximation (npHoMF). See section~\ref{sec:link_importance} for further details.}
   \label{fig:npHoMF}
\end{figure}

We need an expression relating the average degree $\avg{k}$ and the incidence of the epidemics $\rho$ to be able to calculate the approximate values of the \rev{link epidemic} importances in Eqs.~(\ref{seq:importanceA}) to~(\ref{seq:importanceAB}). It can be obtained using the nonperturbative heterogeneous mean field (npHMF) equations in [G\'omez et al., Phys.\ Rev.\ E, 84 (2011) 036105]. In particular, the npHMF equations for the SIS model without one-step reinfections (WOR) read as
\begin{eqnarray}
   0   & = & -\mu \rho_k + (1-\rho_k) (1-q_k)\,,
   \\
   q_k & = & \ds \prod_{k'} (1- \beta \rho_{k'})^{C_{k k'}}\,,
\end{eqnarray}
where $\rho_k$ represents the fraction of infected nodes of degree $k$, $q_k$ the probability that nodes of degree $k$ are not infected by nodes of any other degree $k'$, and $C_{k k'}= k P(k'|k)$ the expected number of links from a node of degree $k$ to nodes of degree $k'$. In the nonperturbative homogeneous mean field (npHoMF) approximation, this reduces to
\begin{eqnarray}
   0   & = & -\mu \rho + (1-\rho) (1-q)\,,
   \\
   q & = & \displaystyle (1- \beta \rho)^{\avg{k}}\,.
\end{eqnarray}
Thus, after some algebra we get
\begin{equation}
   \label{seq:avgk}
   \avg{k} = \frac{\ds\log\left( 1 - \mu \frac{\rho}{1-\rho} \right)}{\log(1 - \beta\rho)}\,.
\end{equation}
An immediate consequence of Eq.~(\ref{seq:avgk}) is that $0 \leqslant \rho \leqslant 1/(1+\mu)$. We can see a plot of the npHoMF relationship between $\rho$ and $\avg{k}$ in the inset of Fig.~\ref{fig:npHoMF}. Note that $\rho$ is an increasing function of $\avg{k}$, thus the larger the average degree, the greater the incidence of the epidemics.

Now, we can substitute Eq.~(\ref{seq:avgk}) for $\avg{k}_A$ and $\avg{k}_B$ into Eqs.~(\ref{seq:importanceA}) to~(\ref{seq:importanceAB}) to obtain approximations of the three different link epidemic importances. The results are presented in Fig.~\ref{fig:npHoMF}. Fixing a certain average degree $\avg{k}_B$ for subnetwork $B$, we consider subnetworks $A$ with $\avg{k}_A \geqslant \avg{k}_B$. Since $\rho$ increases with $\avg{k}$, this is equivalent to fixing $\rho^B$ and consider subnetworks $A$ with $\rho^A \geqslant \rho^B$. We observe that, in all cases, the \rev{link epidemic} importance $I^{A}$ of links in subnetwork $A$ is larger than $I^{AB}$ of the bridge link between subnetworks $A$ and~$B$ ($I^{A} \geqslant I^{AB}$), thus confirming that our epidemic containment strategy driven by link \rev{epidemic} importance does not disconnect the network. This result has been obtained under independence and homogeneous mean field approximations, and for a specific structural configuration of the network. In practice, we observe that the epidemic containment approach based on link epidemic importance is the one considered which better preserves the connectivity of the networks, as shown in Figs.~\ref{fig:synthetic_comps} and~\ref{fig:real_comps}.


\section{Linearization of the ELE model}

The determination of the epidemic threshold from the Epidemic Link Equations (ELE) requires the consideration of states in which the probabilities of having infected nodes are very small, i.e.\ $\condij,\condji,\coupi\ll 1$. Therefore, we may suppose that $\condij,\condji,\coupi \sim O(\varepsilon)$, with $\varepsilon\ll 1$, and in consequence $\coups\sim 1-O(\varepsilon)$. Using these approximations the epidemic link equations become linear in these $O(\varepsilon)$ probabilities, since $O(\varepsilon^2)$ terms should be neglected.

We start with the linearization of the hostility:
\begin{eqnarray}
  \label{seq:hostility}
    h_{ij} & = & \frac{\Phi_{ij}}{\Phi_{ij} + \Theta_{ij}^S} \nn \\
           & = & \frac{\Phi_{ij}}{1 - (\Phi_{ji} + \Theta_{ij}^I)} \nn \\
           & = & \Phi_{ij} \left(1 + \Phi_{ji} + \Theta_{ij}^I + O(\varepsilon^2)\right) \nn \\
           & = & \Phi_{ij} + O(\varepsilon^2)\,,
\end{eqnarray}
where we have used the normalization $\Phi_{ij} + \Phi_{ji} + \Theta_{ij}^I + \Theta_{ij}^S = 1$, and we realize that terms $\Phi_{ij} \Phi_{ji}$ and $\Phi_{ij} \Theta_{ij}^I$ are both $O(\varepsilon^2)$. Substituting hostility in the expression for $q_{ij}$ we get:
\begin{eqnarray}
    q_{ij} & = & \prod_{\substack{r = 1\\r\neq j}}^{N} \left(1-\beta A_{ri} h_{ir}\right) \nn \\
           & = & \prod_{\substack{r = 1\\r\neq j}}^{N} \left(1-\beta A_{ri}\Phi_{ir}+O(\varepsilon^2)\right) \nn \\
           & = & 1 - \beta \sum_{\substack{r = 1\\r\neq j}}^{N} A_{ri}\Phi_{ir} + O(\varepsilon^2) \nn \\
           & = & 1 - \beta \sum_{r = 1}^{N} A_{ri}(1 - \delta_{rj})\Phi_{ir} + O(\varepsilon^2)\,,
\end{eqnarray}
where the Kronecker $\delta_{rj}$ has been introduced to make zero the $j$th term of the sum. Now we are in condition to find the linear approximations of the main ELE model equations. First, the equation for $\Theta_{ij}^I$ becomes:
\begin{eqnarray}
  \label{seq:Theta_mpa}
  \Theta_{ij}^I & = & \Theta^S_{ij}\, (1 - q_{ij})\, (1 - q_{ji}) \nn \\
                &   & \mbox{} + \Phi_{ij}\, (1 - (1 - \beta)q_{ij})\, (1 - \mu) \nn \\
                &   & \mbox{} + \Phi_{ji}\, (1 - \mu)\, (1 - (1 - \beta)q_{ji}) \nn \\
                &   & \mbox{} + \Theta^I_{ij}\, (1 - \mu)^2 \nn \\
                & = & O(\varepsilon^2) \nn \\
                &   & \mbox{} + \Phi_{ij}\, \beta (1 - \mu) + O(\varepsilon^2) \nn \\
                &   & \mbox{} + \Phi_{ji}\, (1 - \mu) \beta + O(\varepsilon^2) \nn \\
                &   & \mbox{} + \Theta^I_{ij}\, (1 - \mu)^2 \nn \\
                & = & \beta (1 - \mu)\, \Phi_{ij} + \beta (1 - \mu)\, \Phi_{ji}
                      + (1 - \mu)^2\, \Theta^I_{ij} + O(\varepsilon^2)\,.
\end{eqnarray}
Note that, for the terms with a factor $\condij$, $\condji$ or $\coupi$, which are $O(\varepsilon)$, we just need to keep $O(1)$ contributions in the rest of the term, thus we may use the approximations $q_{ij} = q_{ji} = 1 + O(\varepsilon)$.

The equation for $\Phi_{ij}$ reads:
\begin{eqnarray}
  \label{seq:Phi_mpa}
  \Phi_{ij} & = & \Theta^S_{ij}\, q_{ij} (1 - q_{ji}) \nn \\
            &   & \mbox{} + \Phi_{ij}\, ((1 - \beta)q_{ij})\, (1 - \mu) \nn \\
            &   & \mbox{} + \Phi_{ji}\, \mu\, (1 - (1 - \beta)q_{ji}) \nn \\
            &   & \mbox{} + \Theta^I_{ij}\, \mu (1 - \mu) \nn \\
            & = & \beta \sum_{r = 1}^{N} A_{rj} (1 - \delta_{ri})\, \Phi_{jr} + O(\varepsilon^2) \nn \\
            &   & \mbox{} + \Phi_{ij}\, (1 - \beta) (1 - \mu) + O(\varepsilon^2) \nn \\
            &   & \mbox{} + \Phi_{ji}\, \mu \beta + O(\varepsilon^2) \nn \\
            &   & \mbox{} + \Theta^I_{ij}\, \mu (1 - \mu) \nn \\
            & = & \beta \sum_{r = 1}^{N} \left(A_{rj} (1 - \delta_{ri}) + \mu \delta_{ri}\right)\, \Phi_{jr} \nn \\
            &   & \mbox{} + \Phi_{ij}\, (1-\beta) (1-\mu) + \Theta^I_{ij}\, \mu (1-\mu) + O(\varepsilon^2) \nn \\
            & = & \beta \sum_{r = 1}^{N} (A_{rj} - (1 - \mu) \delta_{ri})\, \Phi_{jr}
                  + (1 - \beta) (1 - \mu)\, \Phi_{ij} + \mu (1 - \mu)\, \Theta^I_{ij} + O(\varepsilon^2)\,.
\end{eqnarray}
For the last step we have made use of $A_{rj}\delta_{ri} = A_{ij}\delta_{ri} = \delta_{ri}$, since these equations correspond to a link between nodes~$i$ and~$j$, thus we are implicitly assuming that $A_{ij}=1$.

Summarizing, the linearized equations of the ELE model can be expressed as:
\begin{eqnarray}
  \Theta_{ij}^I & = & \beta (1 - \mu)\, \Phi_{ij} + \beta (1 - \mu)\, \Phi_{ji} + (1 - \mu)^2\, \Theta^I_{ij}\,, \\
  \Phi_{ij}     & = & \beta \sum_{r = 1}^{N} (A_{rj} - (1 - \mu) \delta_{ri})\, \Phi_{jr}
                      + (1 - \beta) (1 - \mu)\, \Phi_{ij} + \mu (1 - \mu)\, \Theta^I_{ij}\,.
\end{eqnarray}


\section{Epidemic threshold}

We have shown in Methods that the epidemic threshold is obtained by finding the non-trivial solutions of the system of equations
\begin{equation}
  \label{seq:eigensystem}
  \frac{\mu}{\beta}\epsilon_i = \sum_j B_{ji}\epsilon_j\,,
\end{equation}
where the components of matrix $B$ read
\begin{equation}
  \label{seq:matrix_b}
  B_{ij} = (1 - \Upsilon)A_{ij} - \Upsilon k_i\delta_{ij}\,,
\end{equation}
with the constant $\Upsilon$ being
\begin{equation}
  \Upsilon = \frac{\beta(1-\mu)}{\mu(2-\mu)+2\beta(1-\mu)}\,.
\end{equation}
The non-trivial solutions of Eq.~(\ref{seq:eigensystem}) require $\frac{\mu}{\beta}$ to be an eigenvalue $\Lambda(B)$ of matrix $B$:
\begin{equation}
  \label{seq:eigevalue}
  \Lambda(B) = \frac{\mu}{\beta}
\end{equation}
Unfortunately, for any given fixed value of the recovery rate $\mu$, matrix $B$ also depends on the infection rate $\beta$ through $\Upsilon$, thus Eq.~(\ref{seq:eigevalue}) becomes an implicit equation for $\beta$. Moreover, since we are interested in the onset of the epidemics, we must find the lowest value of the infection probability, $\beta_c$, which satisfies Eq.~(\ref{seq:eigevalue}). At first sight, one would say that we must choose the maximum eigenvalue of $B$ to obtain the lowest value of $\beta$. However, the dependence of $B$ on $\beta$ may rise the question whether a different eigenvalue (e.g., the second largest eigenvalue) could solve the equation at a lower value of $\beta$, since each eigenvalue has a different functional form.

We show in Fig.~\ref{fig:beta_c} that, no matter the functional form of the eigenvalues of $B$ on $\beta$, \rev{if there exists a solution to Eq.~(\ref{seq:eigevalue})}, the curve $\frac{\mu}{\beta}$ always crosses the largest eigenvalue line most-to-the-left than for any other of the eigenvalues, due to the decreasing behavior of $\frac{\mu}{\beta}$. Thus, we can safely say express that
\begin{equation}
  \label{seq:onset}
  \beta_c = \frac{\mu}{\Lambda_{\max}(B)}\,,
\end{equation}
which is the final implicit equation for the epidemic threshold $\beta_c$. In Fig.~\ref{fig:beta_c} we have made use of an Erd\H{o}s-R\'enyi network with 100~nodes, average degree $\avg{k}=6$, and setting $\mu=0.5$, but the previous result is general no matter the network or the parameters.


\section{Data description}

\rev{
Description of the 27~real networks used in Figs.~6 and~\ref{fig:real_comps}, sorted by increasing number of nodes. They have been obtained from the Network Repository (\url{http://networkrepository.com}). Table~\ref{tab:real_nets} provides their main structural characteristics.
\begin{description}
\item [ia-infect-dublin]     Human contact network where nodes represent humans and edges represent proximity (i.e., contacts in the physical world), during the Infectious SocioPatterns event that took place at the Science Gallery in Dublin, Ireland.
\item [soc-wiki-Vote]        Wikipedia voting data from the inception of Wikipedia till January 2008. Nodes represent Wikipedia users and a directed edge from node $i$ to node $j$ represents that user $i$ voted on user $j$.
\item [ca-CSphd]             Genealogy network of PhD's in computer science.
\item [ia-fb-messages]       The Facebook-like Social Network originate from an online community for students at University of California, Irvine. The dataset includes the users that sent or received at least one message.
\item [soc-hamsterster]      Network of the friendship and family links between users of Hamsterster social network.
\item [socfb-USFCA72]        A social friendship network extracted from Facebook consisting of people (nodes) with edges representing friendship ties.
\item [socfb-nips-ego]       A social friendship network extracted from Facebook consisting of people (nodes) with edges representing friendship ties.
\item [socfb-Santa74]        A social friendship network extracted from Facebook consisting of people (nodes) with edges representing friendship ties.
\item [ca-GrQc]              Collaboration network of arXiv General Relativity. Nodes represent scientists, and links coauthorship.
\item [web-spam]             Web Spam Challenge 2008 network.
\item [power-US-Grid]        US Power grid graph.
\item [ca-Erdos992]          Erd\H{o}s collaboration network. Nodes represent scientists, and links coauthorship.
\item [soc-advogato]         Advogato is a social community platform where users can explicitly express weighted trust relationships among themselves. The dataset contains a list of all of the user-to-user links.
\item [p2p-Gnutella08]       Gnutella peer to peer network from August 8 2002.
\item [ia-reality]           Reality mining network data consists of human mobile phone call events between a small set of core users at the MIT whom actually were assigned mobile phones for which all calls were collected. A node represents a person; an edge indicates a phone call or voicemail between two users.
\item [ca-HepTh]             Collaboration network of arXiv High Energy Physics Theory. Nodes represent scientists, and links coauthorship.
\item [soc-anybeat]          Anybeat is an online community, a public gathering place where you can interact with people from around your neighborhood or across the world.
\item [ca-AstroPh]           Collaboration network of arXiv Astrophysics. Nodes represent scientists, and links coauthorship.
\item [ca-CondMat]           Collaboration network of arXiv Condensed Matter. Nodes represent scientists, and links coauthorship.
\item [soc-gplus]            Google+ social network.
\item [tech-as-caida2007]    Internet network at the level of autonomous systems as of 2017. Nodes represent autonomous systems, and there exists a link between them if they have a business agreement for the routing of packets.
\item [ia-email-EU]          The network was generated using email data from a large European research institution. Nodes are users and edges represent email exchanges between two users in both directions.
\item [ia-enron-large]       Network of emails exchanged between senior managers of Enron Corporation, during the period which lead to its bankruptcy.
\item [soc-brightkite]       Brightkite is a location-based social networking service provider where users shared their locations by checking-in. The dataset contains all links among users.
\item [soc-epinions]         Who-trust-whom online social network of the general consumer review site Epinions.com.
\item [soc-slashdot]         A technology-related news website known for its specific user community. The dataset contains friend/foe tags between the users of slashdot.
\item [soc-twitter-follows]  Twitter follower network.
\end{description}
The prefix in the name of each network indicates the cathegory it belongs to, namely: (ia) interaction networks; (soc) social networks; (ca) collaboration networks; (socfb) Facebook networks; (web) we graphs; (power) power networks; (p2p) peer to peer networks; (tech) technological networks.
}

\clearpage

\clearpage

\begin{figure}[!t]
   \centering
   \includegraphics[width=12cm]{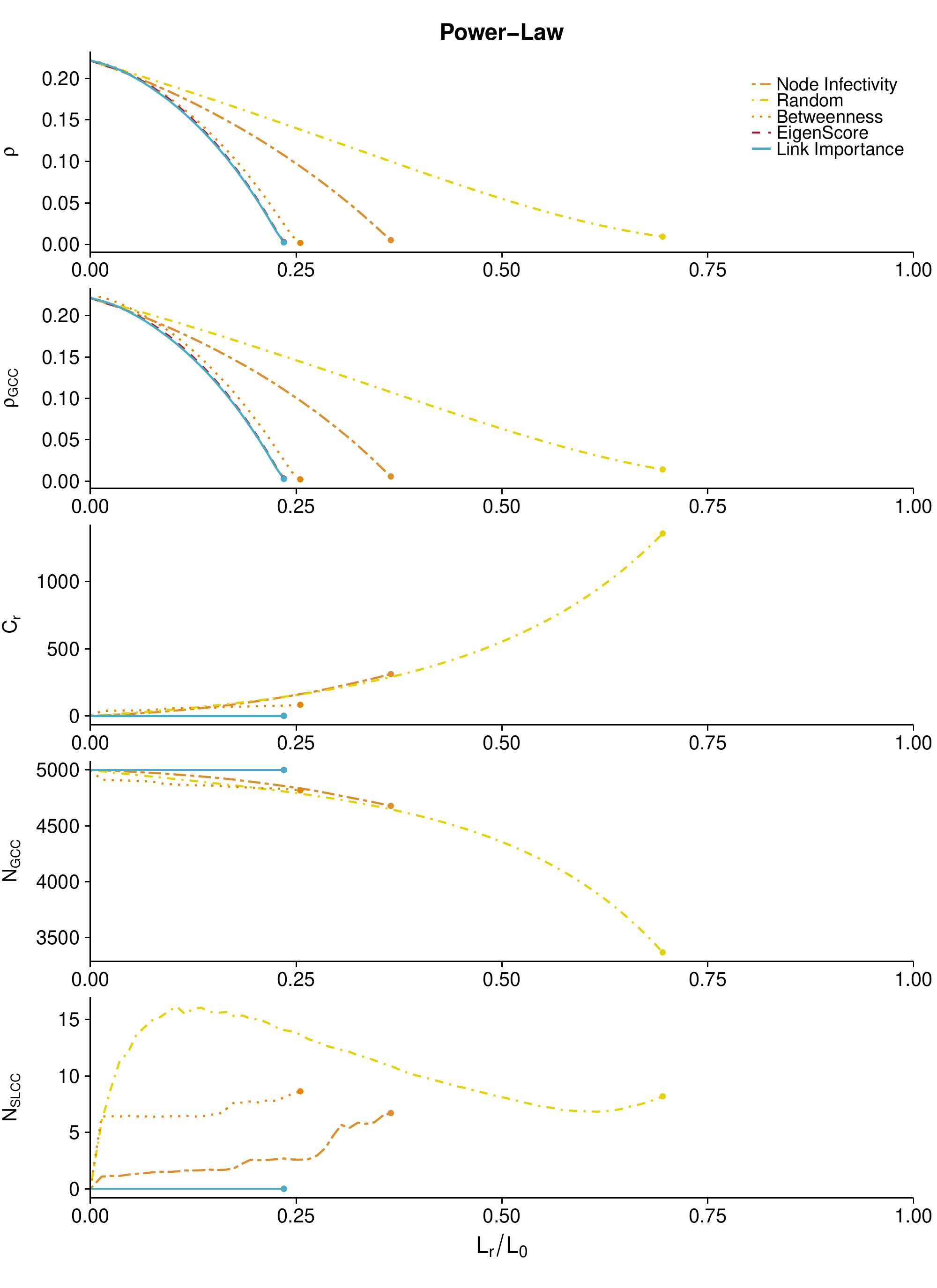}
   \caption{Epidemic containment for a network with 5000~nodes, power-law degree distribution of exponent~3, and average degree $\avg{k}=6$. Five containment strategies are compared: maximum probability of being infected (Node Infectivity), random link selection (Random), maximum edge betweenness (Betweenness), maximum eigenscore (EigenScore), and maximum link \rev{epidemic} importance (Link Importance). In the horizontal axis we represent the fraction of removed links during the containment process ($L_r/L$). We show, from top to bottom: $\rho$, incidence of the epidemics on the network; $\rho_{\mbox{\sz GCC}}$, incidence of the epidemics on the Giant Connected Component; $C_r$, number of connected components; $N_{\mbox{\sz GCC}}$, size of the GCC; $N_{\mbox{\sz SLCC}}$, size of the Second Largest Connected Component. The dots mark the achievement of total containment.}
   \label{fig:info_sf}
\end{figure}

\begin{figure}[!t]
   \centering
   \includegraphics[width=12cm]{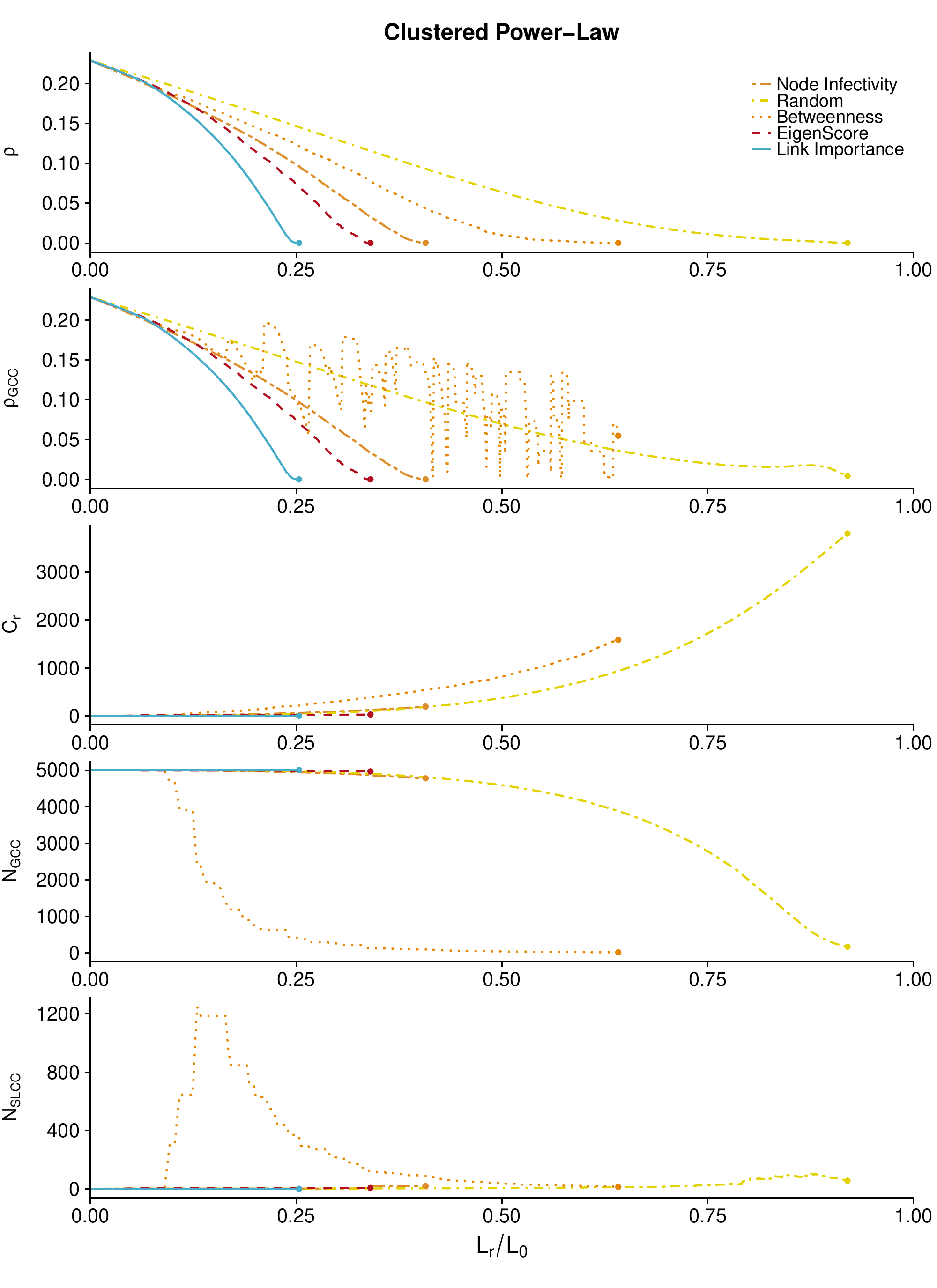}
   \caption{Epidemic containment for a network with 5000~nodes, power-law degree distribution of exponent~3, high clustering coefficient, and average degree $\avg{k}=6$ (see Methods for more details). Five containment strategies are compared: maximum probability of being infected (Node Infectivity), random link selection (Random), maximum edge betweenness (Betweenness), maximum eigenscore (EigenScore), and maximum link \rev{epidemic} importance (Link Importance). In the horizontal axis we represent the fraction of removed links during the containment process ($L_r/L$). We show, from top to bottom: $\rho$, incidence of the epidemics on the network; $\rho_{\mbox{\sz GCC}}$, incidence of the epidemics on the Giant Connected Component; $C_r$, number of connected components; $N_{\mbox{\sz GCC}}$, size of the GCC; $N_{\mbox{\sz SLCC}}$, size of the Second Largest Connected Component. The dots mark the achievement of total containment.}
   \label{fig:info_sfclus}
\end{figure}

\begin{figure}[!t]
   \centering
   \includegraphics[width=12cm]{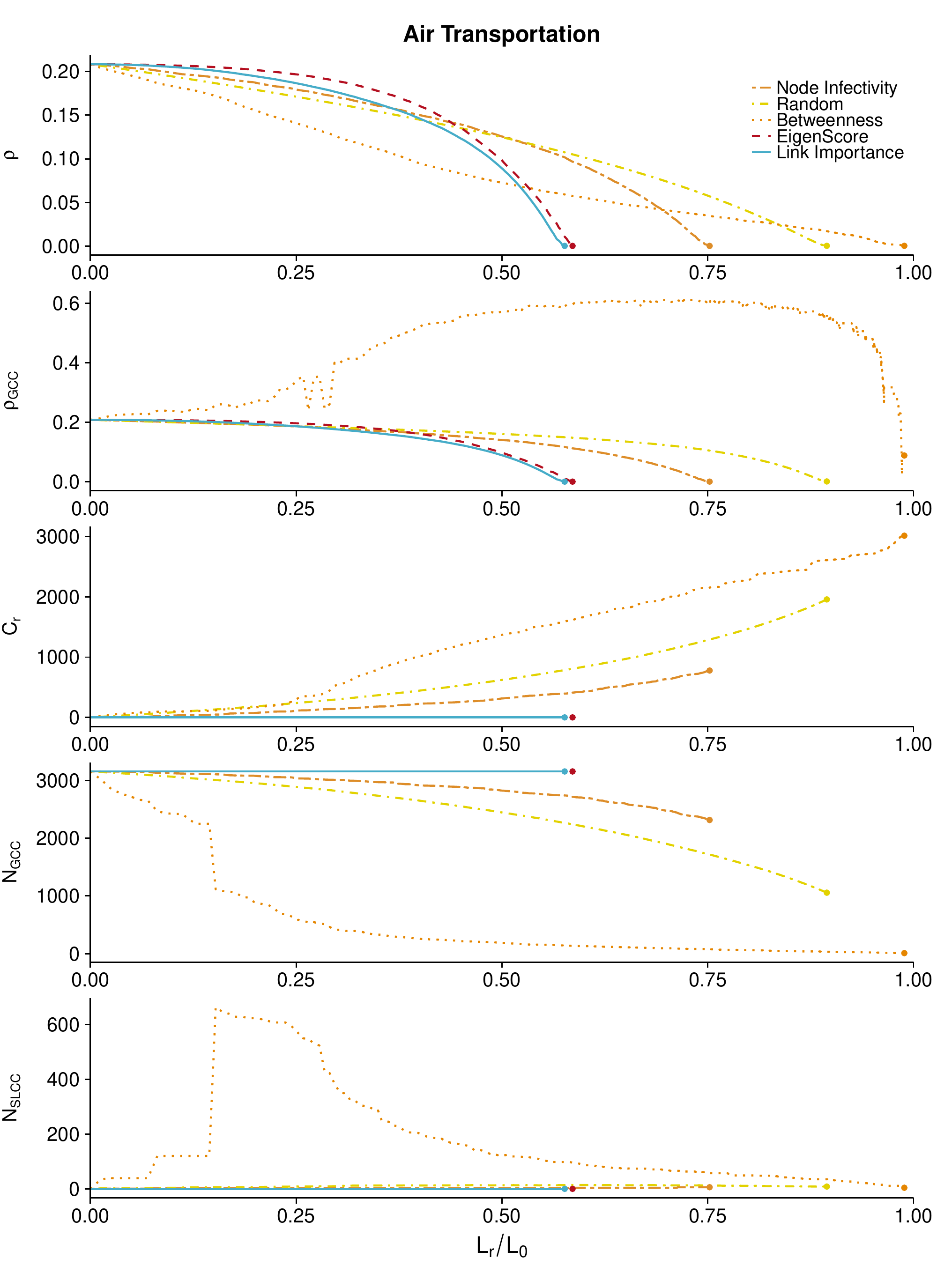}
   \caption{Epidemic containment for the air transportation network (see Methods for more details). Five containment strategies are compared: maximum probability of being infected (Node Infectivity), random link selection (Random), maximum edge betweenness (Betweenness), maximum eigenscore (EigenScore), and maximum link \rev{epidemic} importance (Link Importance). In the horizontal axis we represent the fraction of removed links during the containment process ($L_r/L$). We show, from top to bottom: $\rho$, incidence of the epidemics on the network; $\rho_{\mbox{\sz GCC}}$, incidence of the epidemics on the Giant Connected Component; $C_r$, number of connected components; $N_{\mbox{\sz GCC}}$, size of the GCC; $N_{\mbox{\sz SLCC}}$, size of the Second Largest Connected Component. The dots mark the achievement of total containment.}
   \label{fig:info_air}
\end{figure}

\begin{figure}[!t]
   \centering
   \includegraphics[width=12cm]{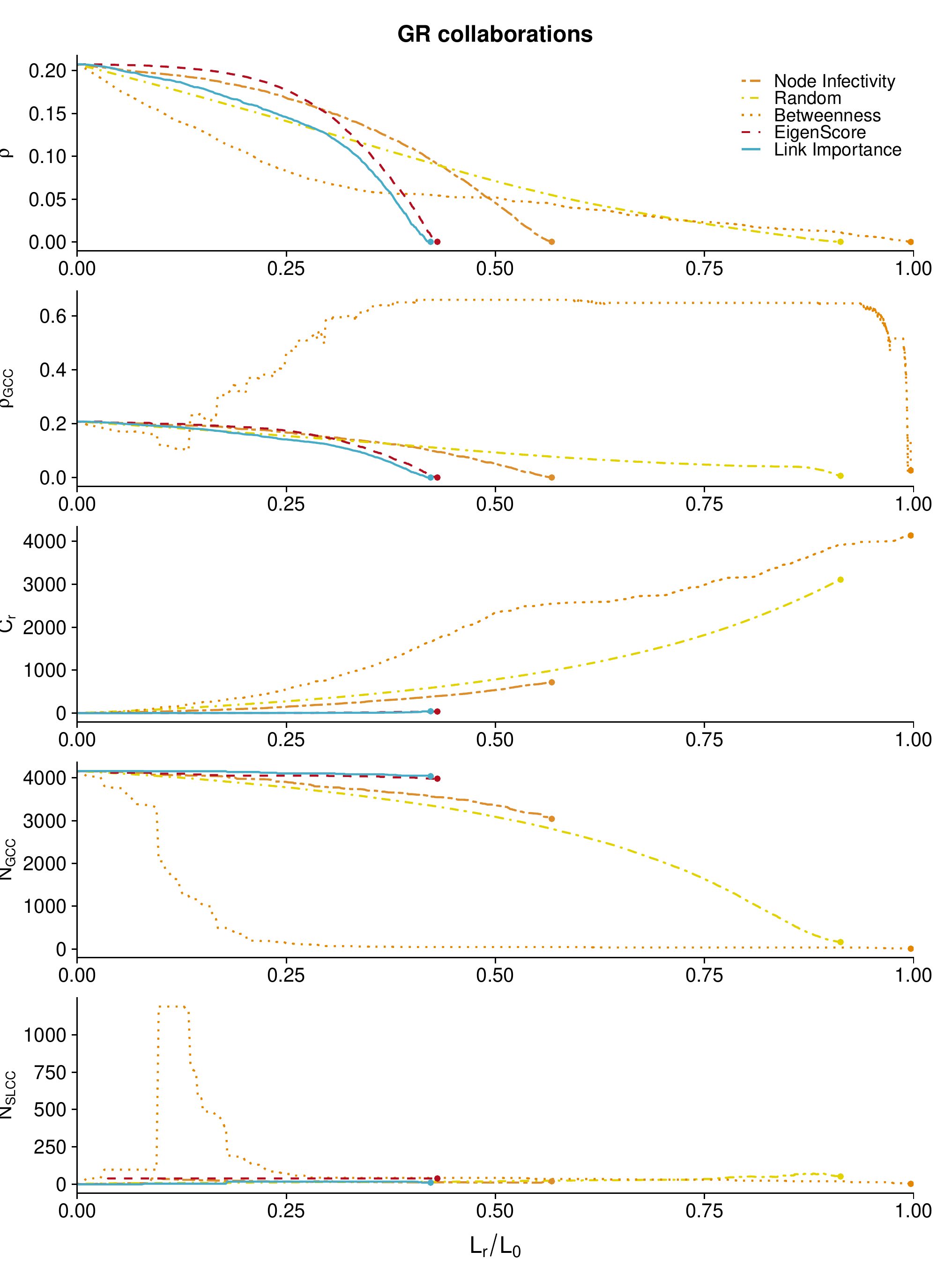}
   \caption{Epidemic containment for the general relativity collaborations network (see Methods for more details). Five containment strategies are compared: maximum probability of being infected (Node Infectivity), random link selection (Random), maximum edge betweenness (Betweenness), maximum eigenscore (EigenScore), and maximum link \rev{epidemic} importance (Link Importance). In the horizontal axis we represent the fraction of removed links during the containment process ($L_r/L$). We show, from top to bottom: $\rho$, incidence of the epidemics on the network; $\rho_{\mbox{\sz GCC}}$, incidence of the epidemics on the Giant Connected Component; $C_r$, number of connected components; $N_{\mbox{\sz GCC}}$, size of the GCC; $N_{\mbox{\sz SLCC}}$, size of the Second Largest Connected Component. The dots mark the achievement of total containment.}
   \label{fig:info_gr}
\end{figure}

\begin{figure}[!t]
   \centering
   \includegraphics[width=12cm]{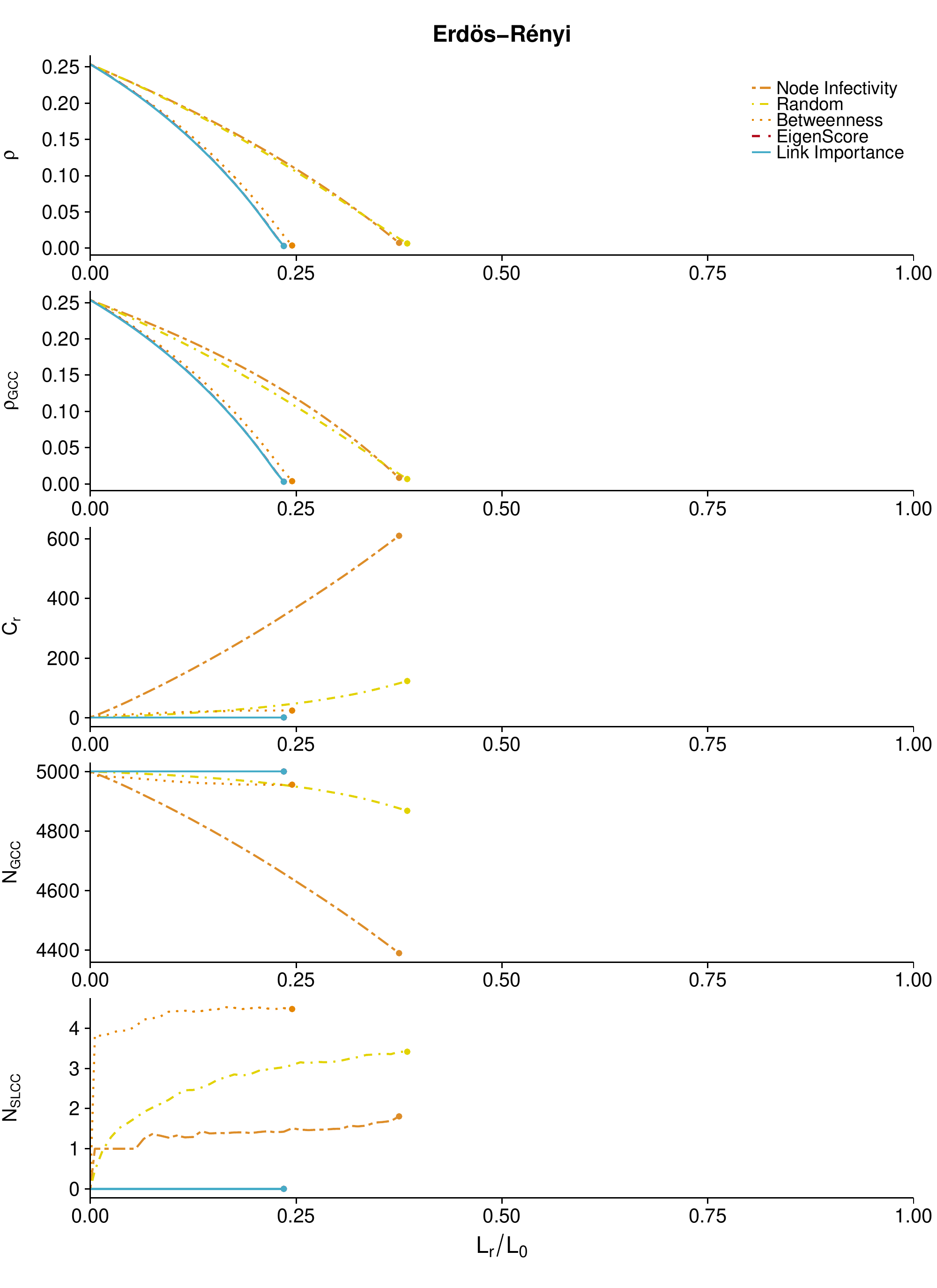}
   \caption{Epidemic containment for an Erd\H{o}s-R\'enyi network with 5000~nodes and average degree $\avg{k}=6$. Five containment strategies are compared: maximum probability of being infected (Node Infectivity), random link selection (Random), maximum edge betweenness (Betweenness), maximum eigenscore (EigenScore), and maximum link \rev{epidemic} importance (Link Importance). In the horizontal axis we represent the fraction of removed links during the containment process ($L_r/L$). We show, from top to bottom: $\rho$, incidence of the epidemics on the network; $\rho_{\mbox{\sz GCC}}$, incidence of the epidemics on the Giant Connected Component; $C_r$, number of connected components; $N_{\mbox{\sz GCC}}$, size of the GCC; $N_{\mbox{\sz SLCC}}$, size of the Second Largest Connected Component. The dots mark the achievement of total containment. We observe that the containment is similar to the one in the power-law network in Fig.~\ref{fig:info_sf} due to the absence of transitivity and modular structure.}
   \label{fig:info_er}
\end{figure}

\begin{figure}[!t]
   \centering
   \includegraphics[width=12cm]{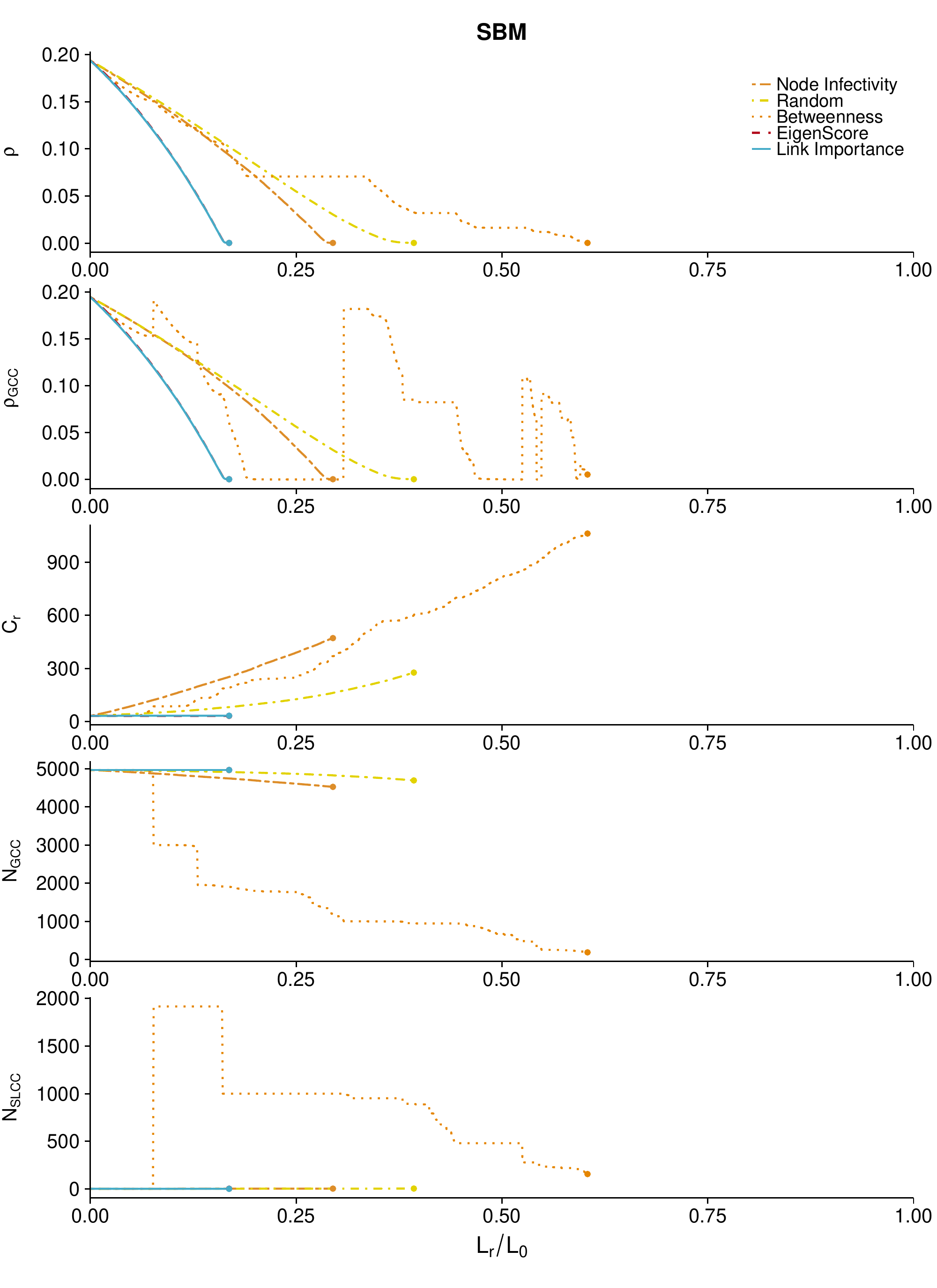}
   \caption{Epidemic containment for a network with 5000~nodes generated with a Stochastic Block Model, with 4~blocks of 250~nodes, 2~blocks of 1000~nodes, and~1 block of 2000~nodes, average degree~5 and mixing probability~0.3. Five containment strategies are compared: maximum probability of being infected (Node Infectivity), random link selection (Random), maximum edge betweenness (Betweenness), maximum eigenscore (EigenScore), and maximum link \rev{epidemic} importance (Link Importance). In the horizontal axis we represent the fraction of removed links during the containment process ($L_r/L$). We show, from top to bottom: $\rho$, incidence of the epidemics on the network; $\rho_{\mbox{\sz GCC}}$, incidence of the epidemics on the Giant Connected Component; $C_r$, number of connected components; $N_{\mbox{\sz GCC}}$, size of the GCC; $N_{\mbox{\sz SLCC}}$, size of the Second Largest Connected Component. The dots mark the achievement of total containment. We observe that our containment strategy based on link \rev{epidemic} importance outperforms all other methods except eigenscore, with similar results, and the large fragmentation induced by the other strategies.}
   \label{fig:info_sbm}
\end{figure}

\begin{figure}[!t]
   \centering
   \includegraphics[width=12cm]{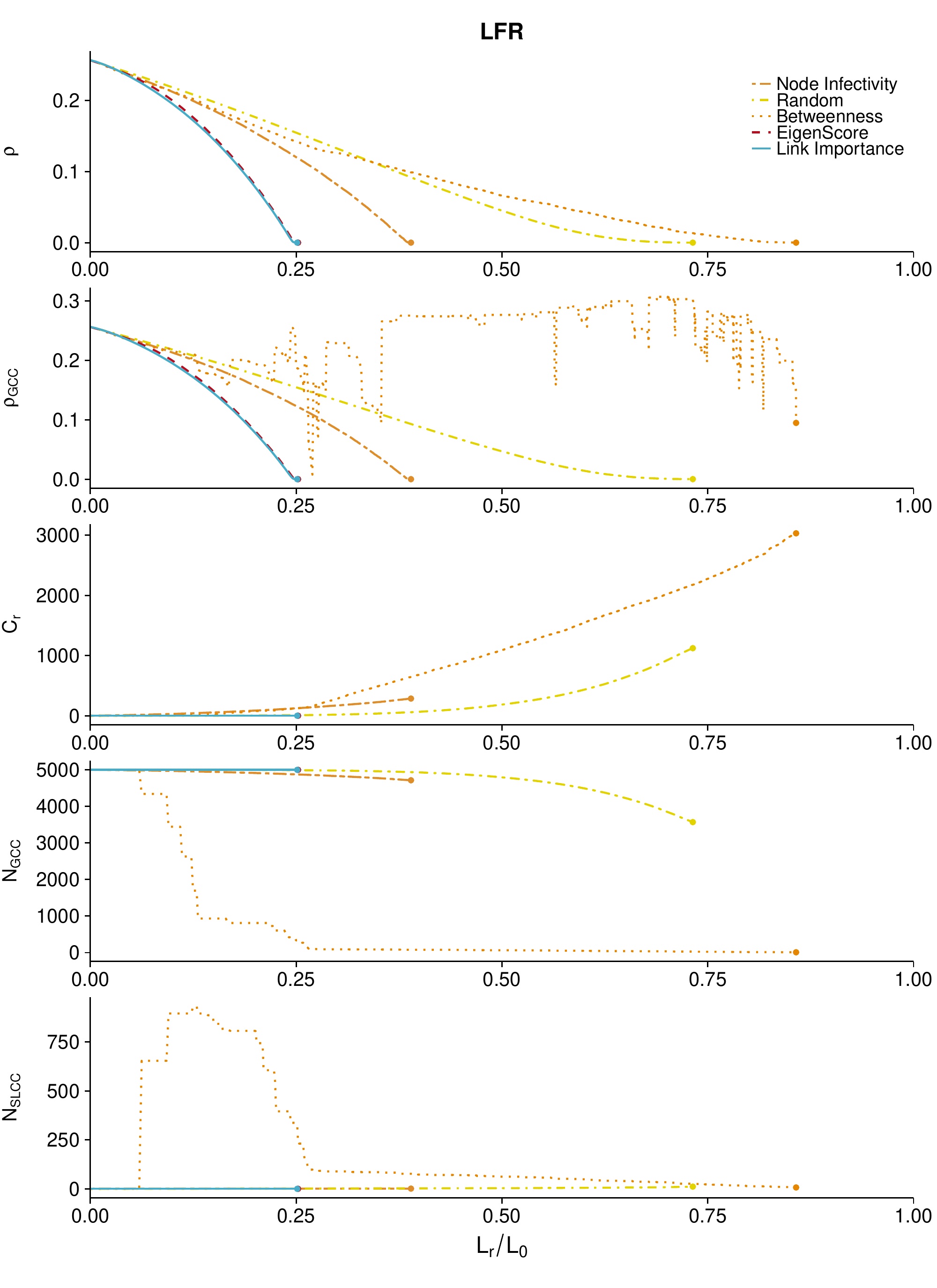}
   \caption{Epidemic containment for a network with 5000~nodes generated using the LFR algorithm~[Lancichinetti et al., Phys.\ Rev.\ E, 78 (2008) 046110], with average degree~6, exponent~3, and mixing probability~0.1. Five containment strategies are compared: maximum probability of being infected (Node Infectivity), random link selection (Random), maximum edge betweenness (Betweenness), maximum eigenscore (EigenScore), and maximum link \rev{epidemic} importance (Link Importance). In the horizontal axis we represent the fraction of removed links during the containment process ($L_r/L$). We show, from top to bottom: $\rho$, incidence of the epidemics on the network; $\rho_{\mbox{\sz GCC}}$, incidence of the epidemics on the Giant Connected Component; $C_r$, number of connected components; $N_{\mbox{\sz GCC}}$, size of the GCC; $N_{\mbox{\sz SLCC}}$, size of the Second Largest Connected Component. The dots mark the achievement of total containment. We observe that our containment strategy based on link \rev{epidemic} importance outperforms all other methods except eigenscore, with similar results, and the large fragmentation induced by the other approaches.}
   \label{fig:info_lfr}
\end{figure}

\begin{figure}[!t]
   \centering
   \includegraphics[width=14cm]{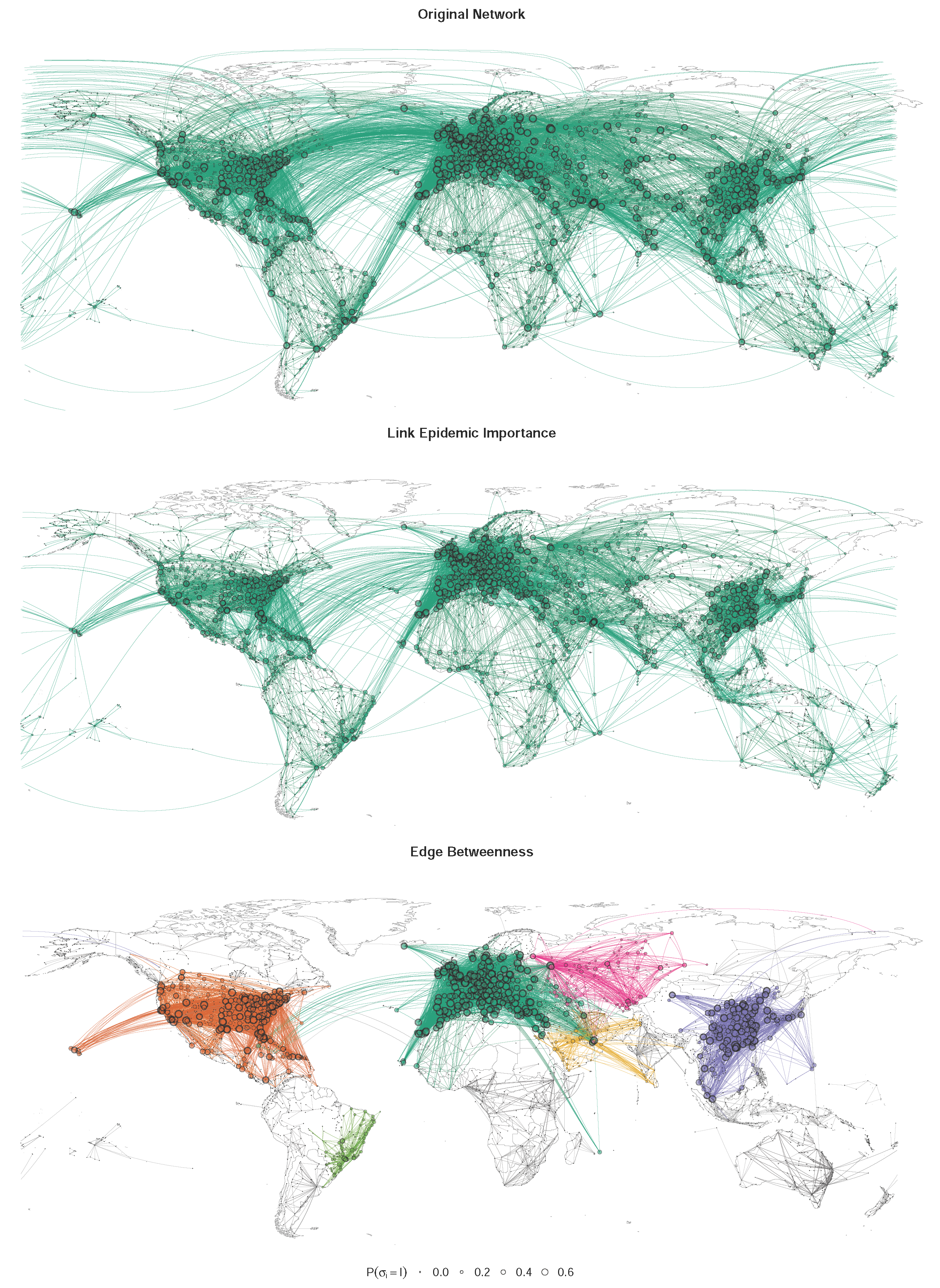}
   \caption{Original air transportation network (top) and the results after a removal of 33.3\% of the links using link \rev{epidemic} importance (middle) and edge betweenness (bottom). Nodes and edges with the same color belong to the same connected component, with subcritical components in gray scale and using darker gray for larger components. The area of the nodes is proportional to their probability of being infected. We have set the epidemic parameters to $\mu=0.5$ and $\beta=0.06$.}
   \label{fig:airports}
\end{figure}

\begin{figure}[!t]
   \centering
   \includegraphics[width=0.95\textwidth]{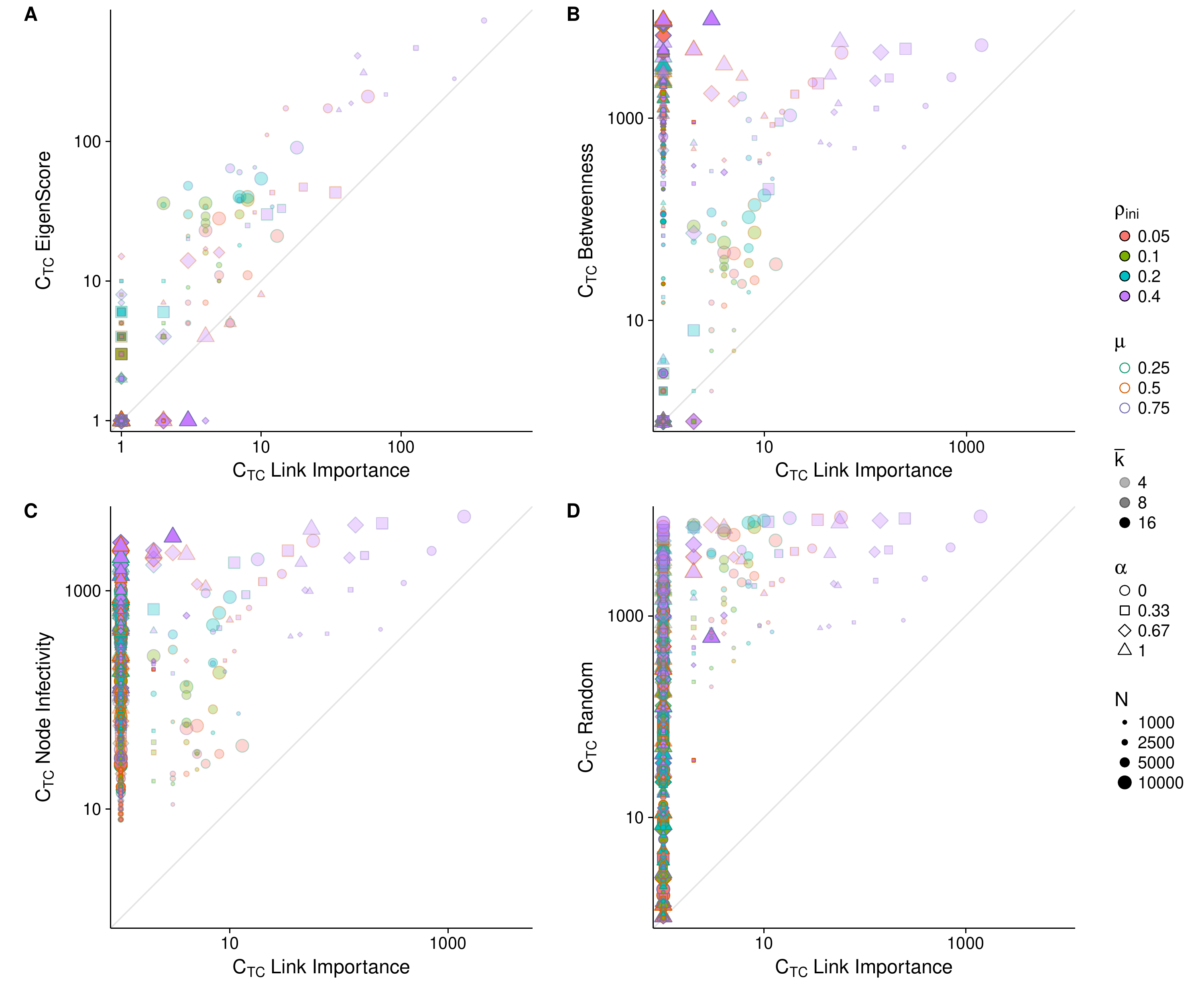}
   \caption{Comparison of the number of connected components after total containment between the \rev{link} epidemic importance strategy and the other four methods, calculated for the synthetic networks and parameters as in Fig.~4.}
   \label{fig:synthetic_comps}
\end{figure}

\begin{figure}[!t]
   \centering
   \includegraphics[width=12cm]{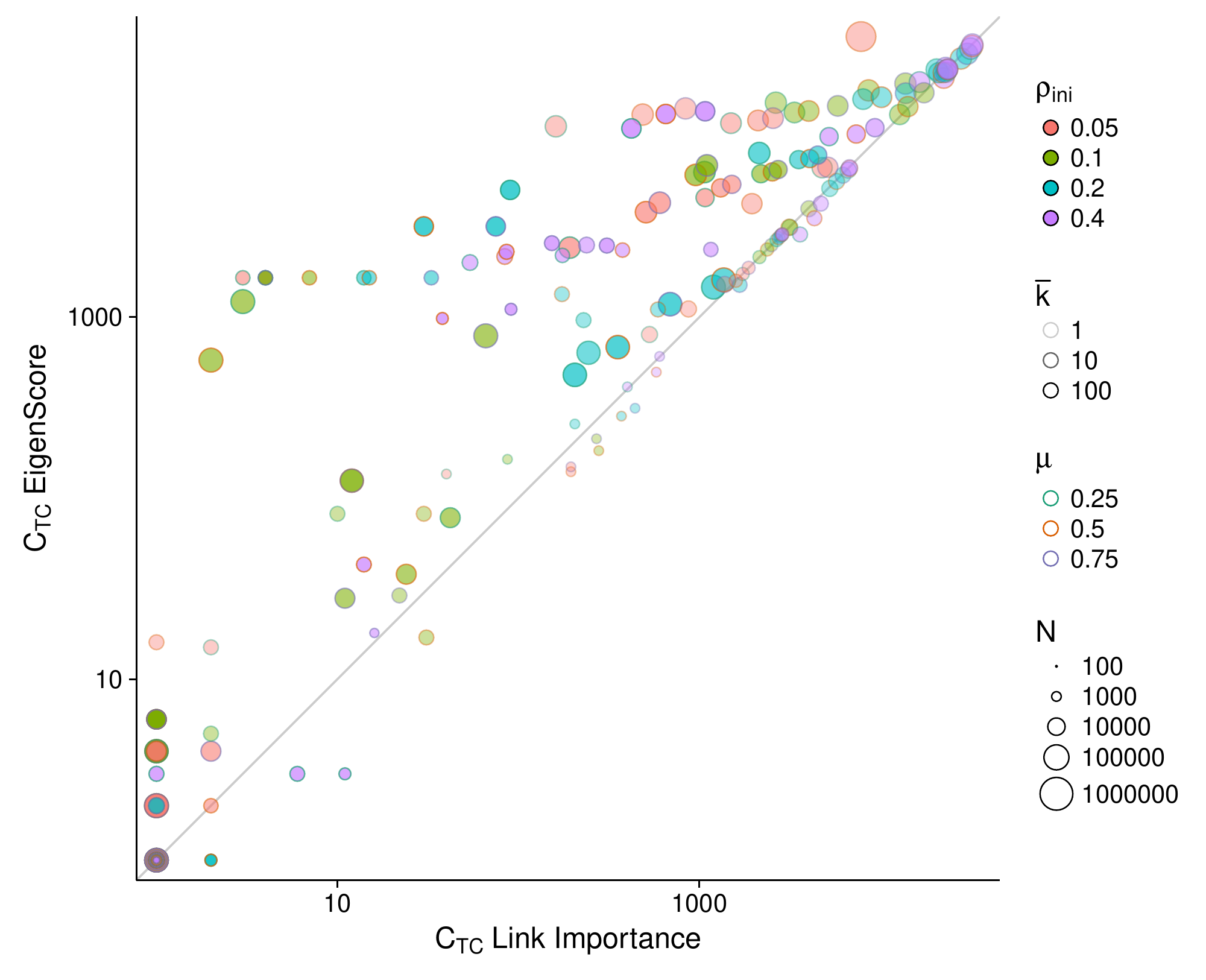}
   \caption{Comparison of the number of connected components after total containment between the \rev{link} epidemic importance and eigenscore strategies, calculated for the real networks and parameters as in Fig.~5.}
   \label{fig:real_comps}
\end{figure}

\begin{figure}[!t]
   \centering
   \includegraphics[width=15cm]{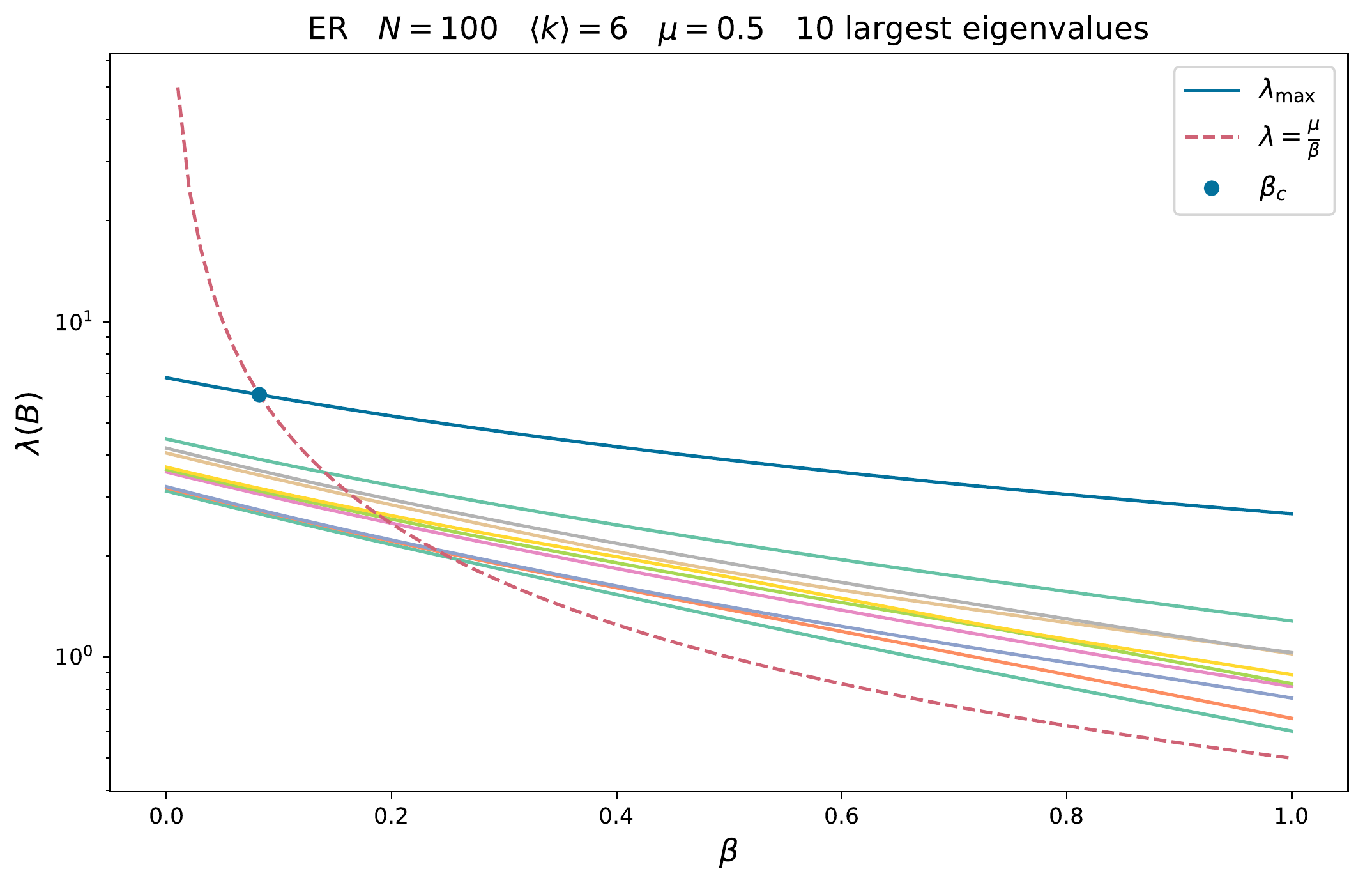}
   \caption{Graphical representation of the determination of the epidemic threshold. The solid lines show the 10~largest eigenvalues of matrix $B$, which depends on the infection probability $\beta$ and the recovery probability $\mu$, for an Erd\H{o}s-R\'enyi network of 100~nodes, average degree $\avg{k}=6$, and setting $\mu=0.5$. For the epidemic threshold, the eigenvalue must be equal to $\frac{\mu}{\beta_c}$, thus it must lay at the intersection between the eigenvalues curves and the dashed line $\frac{\mu}{\beta}$. From those intersections, the one providing the smallest value of $\beta$ is given by the largest eigenvalue of $B$, thus showing that $\beta_c=\frac{\mu}{\Lambda_{\max}(B)}$, which is an implicit equation for the epidemic threshold.}
   \label{fig:beta_c}
\end{figure}

\begin{figure}[!t]
   \centering
   \includegraphics[width=15cm]{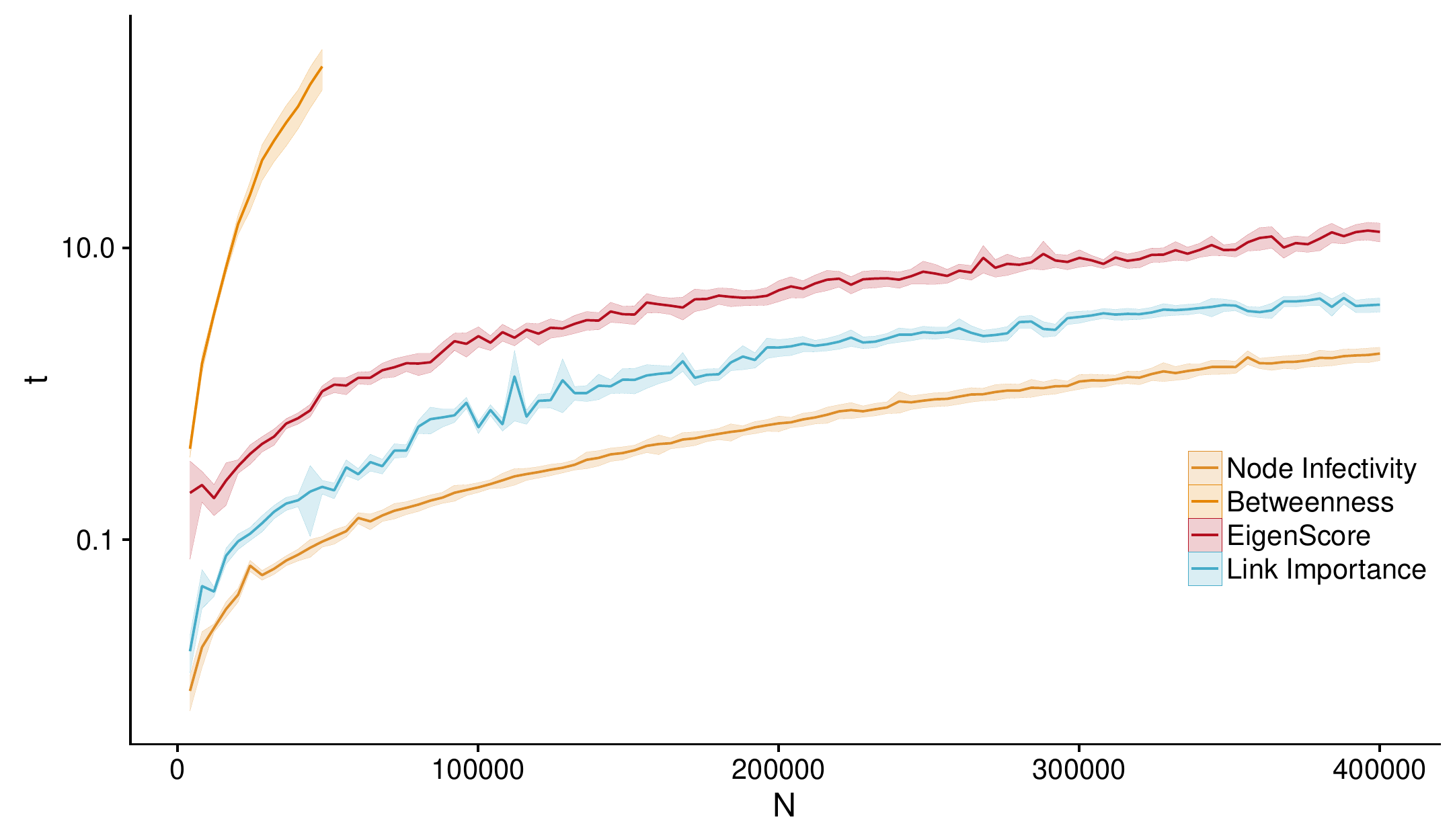}
   \caption{Computational time invested for each method to perform a single ranking and removal for Barab\'asi-Albert networks ranging from~100 to 400,000~nodes, averaged over 36~repetitions. The standard deviations are included as ribbons on the plot.}
   \label{fig:cost}
\end{figure}

\begin{table}[!t]
  \caption{Structural characteristics of the 27~real networks obtained from the Network Repository (\url{http://networkrepository.com}) and used in Fig.~6 and fig.~\ref{fig:real_comps}. They correspond to the largest connected component of the networks with the same name in the repository. The structural descriptors shown are the number of nodes $N$, the number of links $L$, the average degree $\avg{k}$, the clustering coefficient $c$, and the assortativity $r$.}
  \label{tab:real_nets}
  \vspace{15mm}
  \centering
  \begin{tabular}{p{4cm}rrrrr}
  \hline
    Name & \multicolumn{1}{c}{$N$} & \multicolumn{1}{c}{$L$} & \multicolumn{1}{c}{$\avg{k}$}
         & \multicolumn{1}{c}{$c$} & \multicolumn{1}{c}{$r$} \\
  \hline
    ia-infect-dublin     &     410 &   2765 & 13.4878 & 0.1452 &  0.0000 \\
    soc-wiki-Vote        &     889 &   2914 &  6.5557 & 0.0901 & -0.0556 \\
    ca-CSphd             &    1025 &   1043 &  2.0351 & 0.0023 & -0.2532 \\
    ia-fb-messages       &    1266 &   6451 & 10.1912 & 0.6289 &  0.6392 \\
    soc-hamsterster      &    2000 &  16097 & 16.0970 & 0.0268 & -0.3816 \\
    socfb-USFCA72        &    2672 &  65244 & 48.8353 & 0.0037 & -0.3886 \\
    socfb-nips-ego       &    2888 &   2981 &  2.0644 & 0.3178 &  0.2013 \\
    socfb-Santa74        &    3578 & 151747 & 84.8222 & 0.2618 &  0.1253 \\
    ca-GrQc              &    4158 &  13422 &  6.4560 & 0.2294 &  0.0227 \\
    web-spam             &    4767 &  37375 & 15.6807 & 0.0006 & -0.8764 \\
    power-US-Grid        &    4941 &   6594 &  2.6691 & 0.0925 & -0.0952 \\
    ca-Erdos992          &    4991 &   7428 &  2.9766 & 0.0439 & -0.0844 \\
    soc-advogato         &    5054 &  39374 & 15.5813 & 0.0207 &  0.0355 \\
    p2p-Gnutella08       &    6299 &  20776 &  6.5966 & 0.0420 & -0.4531 \\
    ia-reality           &    6809 &   7680 &  2.2558 & 0.1032 &  0.0035 \\
    ca-HepTh             &    8638 &  24806 &  5.7435 & 0.1907 &  0.0917 \\
    soc-anybeat          &   12645 &  49132 &  7.7710 & 0.0024 & -0.6753 \\
    ca-AstroPh           &   17903 & 196972 & 22.0044 & 0.4357 &  0.2258 \\
    ca-CondMat           &   21363 &  91286 &  8.5462 & 0.1273 & -0.0288 \\
    soc-gplus            &   23576 &  39145 &  3.3207 & 0.0073 & -0.1946 \\
    tech-as-caida2007    &   26475 &  53381 &  4.0326 & 0.0260 & -0.0651 \\
    ia-email-EU          &   32430 &  54397 &  3.3547 & 0.0004 & -0.6682 \\
    ia-enron-large       &   33696 & 180811 & 10.7319 & 0.2022 &  0.0706 \\
    soc-brightkite       &   56739 & 212945 &  7.5061 & 0.2811 &  0.2389 \\
    soc-epinions         &   61355 & 494372 & 16.1151 & 0.0217 & -0.1234 \\
    soc-slashdot         &   70068 & 358647 & 10.2371 & 0.0851 & -0.1165 \\
    soc-twitter-follows  &  404719 & 713319 &  3.5250 & 0.1105 &  0.0096 \\
  \hline \\
  \end{tabular}
\end{table}

\end{document}